\documentclass[]{pasj01}
\usepackage{ulem}
\definecolor{gray}{rgb}{0.5, 0.5, 0.5}

\begin{document} 
\Received{}
\Accepted{}

\title{
Follow-up Observations for IceCube-170922A:
Detection of Rapid Near-Infrared Variability and 
Intensive Monitoring of TXS 0506+056
}

%
%
%
\author{Tomoki \textsc{Morokuma}\altaffilmark{1}}
\email{tmorokuma@ioa.s.u-tokyo.ac.jp}
\altaffiltext{1}{Institute of Astronomy, Graduate School of Science, The University of Tokyo, 2-21-1, Osawa, Mitaka, Tokyo 181-0015, Japan}
\author{Yousuke \textsc{Utsumi}\altaffilmark{2}}
\author{Kouji \textsc{Ohta}\altaffilmark{3}}
\author{Masayuki \textsc{Yamanaka}\altaffilmark{4}}
\author{Koji \textsc{Kawabata}\altaffilmark{5}}
\author{Yoshiyuki \textsc{Inoue}\altaffilmark{6,7,8}}
\author{Masaomi \textsc{Tanaka}\altaffilmark{9}}
\author{Michitoshi \textsc{Yoshida}\altaffilmark{10}}
\author{Ryosuke \textsc{Itoh}\altaffilmark{11}}
\author{Mahito \textsc{Sasada}\altaffilmark{5}}
\author{Nozomu \textsc{Tominaga}\altaffilmark{12,7}}
\author{Hiroki \textsc{Mori}\altaffilmark{13}}
\author{Miho \textsc{Kawabata}\altaffilmark{3}}
\author{Tatsuya \textsc{Nakaoka}\altaffilmark{5}}
\author{Maiko \textsc{Chogi}\altaffilmark{13}}
\author{Taisei \textsc{Abe}\altaffilmark{13}}
\author{Ruochen \textsc{Huang}\altaffilmark{13}}
\author{Naoki \textsc{Kawahara}\altaffilmark{13}}
\author{Hiroki \textsc{Kimura}\altaffilmark{13}}
\author{Hiroki \textsc{Nagashima}\altaffilmark{13}}
\author{Kengo \textsc{Takagi}\altaffilmark{13}}
\author{Yuina \textsc{Yamazaki}\altaffilmark{13}}
\author{Wei \textsc{Liu}\altaffilmark{14}}
\author{Ryou \textsc{Ohsawa}\altaffilmark{1}}
\author{Shigeyuki \textsc{Sako}\altaffilmark{1}}
\author{Katsuhiro L. \textsc{Murata}\altaffilmark{15}}
\author{Kumiko \textsc{Morihana}\altaffilmark{16,17}}
\author{Christina K. \textsc{Gilligan}\altaffilmark{18}}
\author{Keisuke \textsc{Isogai}\altaffilmark{4}}
\author{Mariko \textsc{Kimura}\altaffilmark{19}}
\author{Yasuyuki \textsc{Wakamatsu}\altaffilmark{3}}
\author{Ryuhei \textsc{Ohnishi}\altaffilmark{3}}
\author{Masaki \textsc{Takayama}\altaffilmark{20}}
\author{Satoshi \textsc{Honda}\altaffilmark{20}}
\author{Yoshiki \textsc{Matsuoka}\altaffilmark{21}}
\author{Takuji \textsc{Yamashita}\altaffilmark{22,21}}
\author{Shigehiro \textsc{Nagataki}\altaffilmark{23,6}}
\author{Yasuyuki T. \textsc{Tanaka}\altaffilmark{5}}
\altaffiltext{2}{Kavli Institute for Particle Astrophysics and Cosmology, SLAC National Accelerator Laboratory, Stanford University, Menlo Park, CA 94025, U.S.A.}
\altaffiltext{3}{Department of Astronomy, Graduate School of Science, Kyoto University, Sakyo-ku, Kyoto 606-8502, Japan}
\altaffiltext{4}{Okayama Observatory, Kyoto University, 3037-5 Honjo, Kamogata-cho, Asakuchi, Okayama 719-0232, Japan}
\altaffiltext{5}{Hiroshima Astrophysical Science Center, Hiroshima University, Kagamiyama 1-3-1, Higashi-Hiroshima, Hiroshima 739-8526, Japan}
\altaffiltext{6}{Interdisciplinary Theoretical \& Mathematical Science Program (iTHEMS), RIKEN, 2-1 Hirosawa, Saitama 351-0198, Japan}
\altaffiltext{7}{Kavli Institute for the Physics and Mathematics of the Universe (WPI), UTIAS, The University of Tokyo, Kashiwa, Chiba 277-8583, Japan}
\altaffiltext{8}{Department of Earth and Space Science, Graduate School of Science, Osaka University, Toyonaka, Osaka 560-0043, Japan}
\altaffiltext{9}{Astronomical Institute, Tohoku University, Aramaki, Aoba-ku, Sendai, 980-8578, Japan}
\altaffiltext{10}{Subaru Telescope, National Astronomical Observatory of Japan, National Institutes of Natural Sciences, 650 North A'ohoku Place, Hilo, HI 96720, USA}
\altaffiltext{11}{Bisei Astronomical Observatory, 1723-70 Okura, Bisei, Ibara, Okayama 714-1411, Japan}
\altaffiltext{12}{Department of Physics, Faculty of Science and Engineering, Konan University, 8-9-1 Okamoto, Kobe, Hyogo 658-8501, Japan}
\altaffiltext{13}{Graduate School of Science, Hiroshima University, 1-3-1 Kagamiyama, Higashi-Hiroshima, Hiroshima 739-8526, Japan}
\altaffiltext{14}{Purple Mountain Observatory, Chinese Academy of Sciences, No.~10 Yuanhua Road, Qixia District, Nanjing, 210023, China}
\altaffiltext{15}{Department of Physics, Tokyo Institute of Technology, 2-12-1 Ookayama, Meguro-ku, Tokyo 152-8551, Japan}
\altaffiltext{16}{Department of Astrophysics, Nagoya University, Chikusa-ku, Nagoya 464-8602, Japan}
\altaffiltext{17}{Institute of Liberal Arts and Sciences, Nagoya University, Furo-cho, Chikusa-ku, Nagoya, 464-8602, Japan}
\altaffiltext{18}{Department of Physics and Astronomy, Dartmouth College, Hanover, NH 03784, U.S.A.}
\altaffiltext{19}{Extreme Natural Phenomena RIKEN Hakubi Research Team, Cluster for Pioneering Research, RIKEN, 2-1 Hirosawa, Wako, Saitama 351-0198, Japan}
\altaffiltext{20}{Nishi-Harima Astronomical Observatory, Center for Astronomy, University of Hyogo, 407-2 Nishigaichi, Sayo-cho, Sayo, Hyogo 679- 5313, Japan}
\altaffiltext{21}{Research Center for Space and Cosmic Evolution, Ehime University, 2-5 Bunkyo-cho, Matsuyama, Ehime 790-8577, Japan}
\altaffiltext{22}{National Astronomical Observatory of Japan, National Institutes of Natural Sciences, 2-21-1, Osawa, Mitaka, Tokyo 181-8588, Japan}
\altaffiltext{23}{Astrophysical Big Bang Laboratory (ABBL), RIKEN Cluster for Pioneering Research, RIKEN, 2-1 Hirosawa, Wako, Saitama 351-0198, Japan}



\maketitle

\newcommand{\thisevent}{IceCube-170922A}
\newcommand{\target}{TXS~0506+056}
\newcommand{\kyototel}{the 0.4-m Kyoto University telescope}
\newcommand{\tomoe}{Tomo-e~Gozen}
\newcommand{\varidaily}{1.0}
\newcommand{\numofbros}{7}
\newcommand{\itohbrospaper}{Itoh et al., submitted}
\newcommand{\morokumasnpaper}{Morokuma et al., in prep.}

\begin{abstract}

We present our follow-up observations to search for an electromagnetic counterpart 
of the IceCube high-energy neutrino, \thisevent. 
Monitoring observations of a likely counterpart, \target, are also described. 
First, we quickly took optical and near-infrared images 
of \numofbros~flat-spectrum radio sources 
within the IceCube error region right after the neutrino detection 
and found a rapid flux decline of \target\ in Kanata/HONIR $J$-band data. 
Motivated by this discovery, intensive follow-up observations of \target\  are continuously done, 
including our monitoring imaging observations, spectroscopic observations, and polarimetric observations 
in optical and near-infrared wavelengths. 
\target\ shows a large amplitude ($\sim\varidaily$~mag) variability 
in a time scale of several days or longer, 
although no significant variability is detected in a time scale of a day or shorter. 
\target\ also shows a bluer-when-brighter trend in optical and near-infrared wavelengths. 
Structure functions of variabilities are examined and indicate that 
\target\ is not a special blazar in terms of optical variability. 
Polarization measurement results of \target\ are also discussed. 
\end{abstract}

\section{Introduction}\label{sec:sec_intro}

Recent detections of high-energy, TeV-PeV, neutrinos realized by the IceCube experiment \citep{aartsen2014} 
have made more exciting electromagnetic identification of the neutrino sources. 
Such high-energy neutrinos are produced by decay of charged pions which are created
through cosmic-ray interactions with radiation ($p\gamma$; \cite{winter2013})
or gas ($pp$; \cite{murase2013}). 
Therefore, 
detection of high-energy neutrino is a smoking-gun evidence of existence of high-energy protons (cosmic rays). 
Observations of high-energy neutrinos 
provide unique information about cosmic-ray acceleration mechanisms 
and their acceleration sites, if their origins are identified.

In \citet{icecube2015}, detections of 54~neutrino events by the IceCube experiment in total 
are reported from the 4-year data.
Arrival directions of the 54~IceCube events are consistent with being 
isotropic and do not show any clustering in the Galactic plane \citep{icecube2015}, 
indicating the neutrino sources would be extragalactic \citep{aartsen2014}. 
Neutrinos can travel cosmological distances without being deflected by cosmic magnetic field 
nor absorbed by photon field. 
On the other hand, contribution of high-redshift sources represents 
only a small fraction of the total observed flux due to the redshift dilution. 
Thus, the competition between the neutrino's penetrating power and 
the redshift dilution makes emissions from sources 
at redshift of $z\sim1-2$ dominant to the IceCube neutrino 
for single-neutrino ({\it singlet})
events (\cite{kotera2010}; \cite{icecube2017_multiplet}). 
The IceCube collaboration started issuing 
real-time alerts in April 2016 \citep{aartsen2017realtimealert} \sout{and} 
and the number of the neutrino detection is increasing. 

Many hypotheses for the origin of the high energy neutrinos have been proposed, 
including 
blazars \citep{mannheim1995,mucke2003,becker2005}, 
starburst galaxies \citep{loeb2006,bechtol2017}, 
Type~IIn supernovae \citep{murase2011,aartsen2015sn2n}, 
choked-jet supernovae \citep{razzaque2004,senno2016}, 
gamma-ray bursts (GRBs; \cite{waxman1997,aartsen2017grb}), 
tidal disruption events (TDEs; \cite{senno2017}), 
active galactic nuclei (AGN) core \citep{eichler1979,inoue2019}. 

These theories have been tried to be assessed observationally and some observational 
results set constraints on the theories as described in the 
this paragraph. 
\citet{kadler2016} showed that 
a high fluence GeV blazar, PKS~1424-418, is a possible origin for 
one of the high-energy starting events (HESEs), HESE-35, 
based on the temporal and positional coincidence 
between the neutrino detection and 
the $\gamma$-ray flare of the blazar. 
This blazar shows broad CIV, CIII], and Mg~II lines in its optical spectrum and 
is classified as a flat-spectrum radio quasar (FSRQ) at $z=1.522$ \citep{white1988}. 
On the other hand, in general, 
there is not a good spatial 
correlation between the neutrino detections and Fermi $\gamma$-ray blazars, 
indicating that contribution of the Fermi $\gamma$-ray blazars 
to the diffuse neutrino flux is 
as small as $<27$\% \citep{aartsen2017blazar}. 
No other origin candidates have been strongly supported 
as a significantly contributing source to the neutrino background 
\citep{bechtol2017,senno2016,senno2017}.  
Very recently, 
a radio-emitting TDE at $z=0.051$ 
discovered on April 1, 2019, 
is claimed to be an associated source to 
the neutrino event IceCube-191001A \citep{stein2020}. 

On 22 September 2017 at 20:54:30.43 (MJD=58018.871), 
the IceCube experiment 
detected a track event of a high-energy neutrino
($\sim290$~TeV; \cite{icecube2018_science1}), 
\thisevent, and 
\citet{kopper2017icecubegcn} reported it 
via 
The Gamma-ray Coordinates Network (GCN) Circular. 
The direction is constrained to be 
R.A.$=77.43^{+0.95}_{-0.65}$ and Dec.$=+5.72^{+0.50}_{-0.30}$ degrees 
in J2000 equinox (90\% confidence region). 
\target, a BL~Lac blazar within the error region, 
was pointed out to be a good candidate of the counterpart 
(\cite{tanaka2017_IC170922a}; see also \S\ref{sec:sec_discnirvari}) 
and observed with many telescopes over a wide wavelength range 
after that \citep{icecube2018_science1}. 
During the intensive follow-up observations, 
the redshift of \target\ was successfully determined to be $z=0.3365$ \citep{paiano2018}. 
In addition, \target\ was detected 
in high-energy $\gamma$-ray with 
Large Area Telescope (LAT) on the Fermi satellite (\cite{tanaka2017_IC170922a}; \cite{icecube2018_science1}), 
MAGIC telescope (\cite{icecube2018_science1}; \cite{ansoldi2018}), 
and AGILE $\gamma$-ray telescope \citep{lucarelli2019}, independently, 
which strength the coincidence between \target\ and the neutrino source.

\target\ is a blazar 
registered in a catalog of 
Texas Survey of Radio Sources 
\citep{douglas1996} and 
one of the highest energy $\gamma$-ray emitting blazars
among blazars detected by the Energetic Gamma Ray Experiment Telescope~(EGRET) 
$\gamma$-ray (30~MeV-30~GeV) satellite \citep{dingus2001}. 
The radio-to-gamma-ray spectral energy distribution (SED; \cite{icecube2018_science1}) in combination with 
its featureless spectra (\cite{paiano2018}; \cite{icecube2018_science1}) 
and peak frequency of $10^{14.5\pm0.25}$~Hz \citep{padovani2019} 
indicates that \target\ belongs to 
intermediate synchrotron peaked 
($10^{14}<\nu_{S}<10^{15}$~Hz) BL~Lac objects 
(ISPs or IBLs; \cite{padovani2018}) 
although \citet{padovani2019} claim that \target\ is a masquerading BL~Lac. 
The bolometric luminosity 
$L_{\rm{bol}}$ 
is estimated to be a few $\times10^{45}$~erg~s$^{-1}$ \citep{padovani2019}, 
which is roughly consistent with being an ISP
by following the so-called blazar sequence 
\citep{fossati1998,kubo1998,ghisellini2017}. 

In SEDs of BL~Lacs, 
non-thermal emission from a relativistic jet dominates thermal emission 
from an accretion disk 
in rest-frame UV-optical wavelengths and 
from a dusty torus in rest-frame near-infrared (NIR) wavelengths, 
as well as its host galaxy. 
Temporal flux (luminosity) variability of blazars is 
sometimes
explained by a shock-in-jet model \citep{marscher1985} and 
useful to see a possible link between 
neutrino emission, probably from a relativistic jet, 
and electromagnetic emission activities. 
To assess if or not \target\ is a special blazar and September 22, 2017 is a special timing, 
it is worth examining variability properties of \target. 
Blazars, in general, show large and rapid variability in optical wavelengths \citep{bauer2009blazar}, 
which could be a key to understanding relationship with neutrino emission. 
Although it is dependent on blazar types, 
intranight variability is significantly detected for 
several tens of percent of blazars 
and its duty cycles are also several tens of percent 
(\cite{paliya2017}; \cite{bachev2017}; \cite{gaur2017}; \cite{rani2011}). 
Thus, short time scale variability is also worth being examined. 

Strong polarization in optical wavelengths 
is one of the unique characteristics of BL~Lacs 
(e.g., \cite{mead1990,falomo2014,ikejiri2011,itoh2016}), 
which supports an idea that synchrotron emission dominates optical emission 
of BL~Lacs. 
Temporal changes of polarization degrees and angles have been observed for many blazars \citep{itoh2016,jermak2016} 
and these observables are keys to understanding structure and physics in relativistic jets of blazars 
\citep{brand1985,visvanathan1998}. 

The structure of this paper is as follows. 
In section \ref{sec:sec_followup}, 
we describe our follow-up observations after the alert of the IceCube-170922A, 
including imaging and spectroscopic observations in the optical and near-infrared wavelengths. 
We describe the observational results and its discussion in section \ref{sec:sec_resultsdiscussion}. 
The neutrino direction and \target\ are 
inside the Pan-STARRS1 (PS1; \cite{chambers2016}) footprint 
and 
outside the Sloan Digital Sky Survey (SDSS; \cite{york2000}) footprint. 
Cosmological parameters used in this paper are 
$\Omega_{M}=0.3, \Omega_{\Lambda}=0.7, H_{0}=70$~km~s$^{-1}$~Mpc$^{-1}$. 
All the observing times are specified in UT. 

\section{Follow-Up Observations and Data Reduction}\label{sec:sec_followup}

\begin{figure*}
 \begin{center}
	\includegraphics[width=145mm,bb=0 0 981 757]{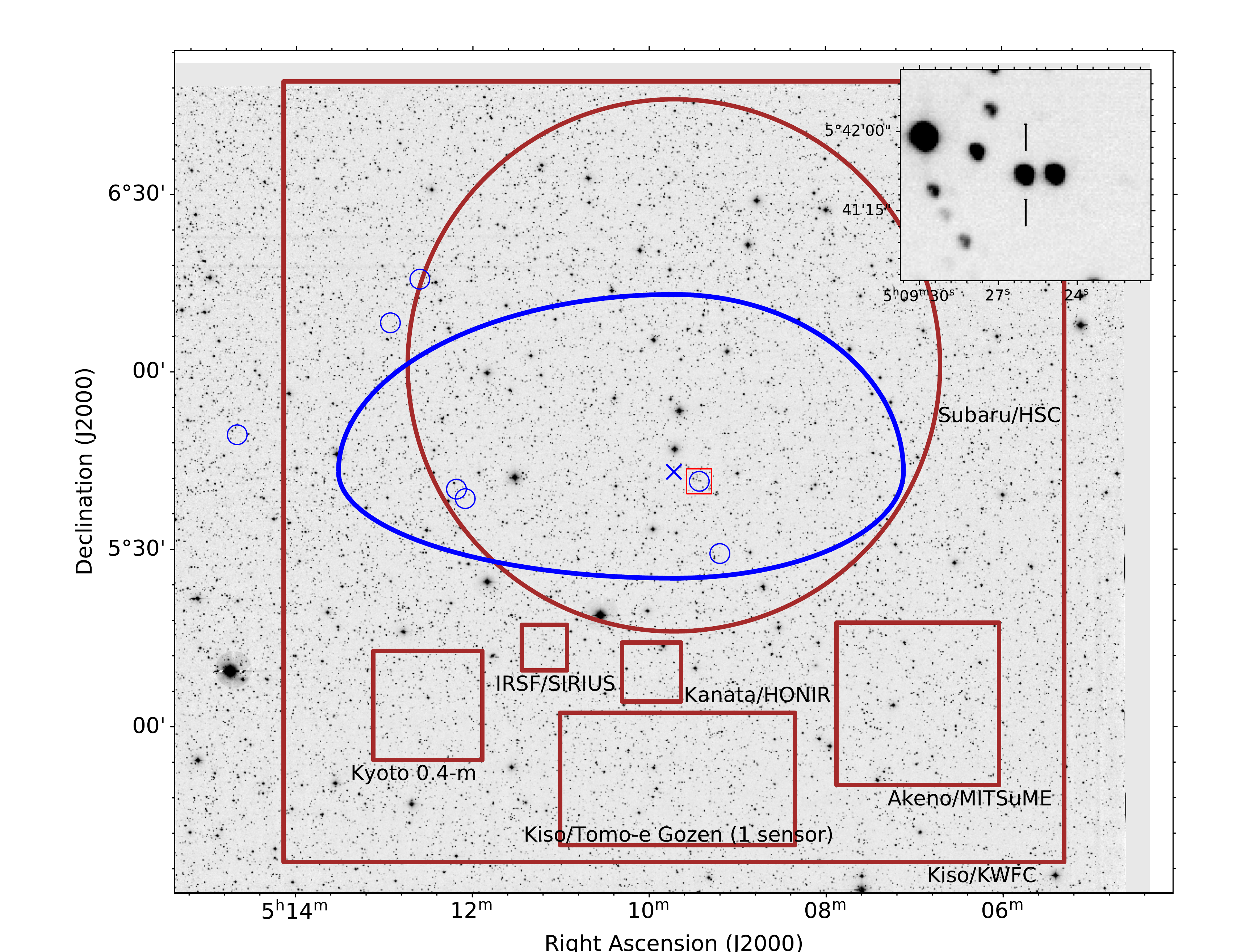}
 \end{center}
\caption{
Coadded $r$-band image of the field taken with KWFC. 
North is up and east is left. 
The IceCube 90\% containment region \citep{icecube2018_science1} is shown in blue thick ellipse and 
its center is indicated as a blue cross. 
The seven flat-spectrum radio sources in the preliminary BROS catalog 
which we observed first are indicated in blue small circles 
and \target\ region is marked as a red rectangle, close to the blue cross. 
Zoom-up $1\times1$~arcmin$^2$ view around \target\ is also shown in the top right inset. 
The field-of-views of the optical and NIR instruments used in this paper are shown in brown. 
For \tomoe, the field-of-view of one sensor is shown. 
}\label{fig:fig_imgfield}
\end{figure*}

We started intensive follow-up observations in optical and NIR wavelengths 
right after receiving the realtime alert of the event. 
In this section, 
we summarize our strategy in general to identify an electromagnetic counterpart 
of an IceCube neutrino-emitting source (\S\ref{sec:sec_searchstrategy}), 
quick observations to search for a counterpart candidate 
of the \thisevent\ 
(\S\ref{sec:sec_optimg}), 
and follow-up observations after a likely counterpart, \target, was identified, 
including monitoring imaging 
and spectroscopic observations 
(\S\ref{sec:sec_optimg},\S\ref{sec:sec_optspec}). 
Polarimetric observations for \target\ are also described (\S\ref{sec:sec_optnirpol}). 

\subsection{Strategy for IceCube-170922A Electromagnetic Counterpart}\label{sec:sec_searchstrategy}

Since the realtime alert system of IceCube was started in April 2016 \citep{aartsen2017realtimealert}, 
we have organized a strategic optical and NIR follow-up observing group 
utilizing the Optical and Infrared Synergetic Telescopes for Education and Research (OISTER; Yamanaka et al. in prep.) 
and other Japanese facilities. 
Considering multiple possibilities for 
transient high-energy neutrino sources, 
we adopt multiple observing strategies using our optical and NIR facilities. 

In order to test the blazar scenario,
we first select blazar candidates 
with flat radio spectra 
catalogued in the BROS catalog 
\citep{itoh2018bros}
The BROS catalog 
collects flat-spectrum radio sources selected using 
a combined catalog \citep{degasperin2018} of 
NRAO Very Large Array Sky survey (NVSS; \cite{condon1998}; 1.4~GHz) and 
TIFR GMRT Sky Survey (TGSS; \cite{intema2017}; 150~MHz) catalogs. 
The final criterion on the radio spectral slope to select flat-spectrum objects is 
$\alpha>-0.6$ where $f_{\nu}\propto\nu^{\alpha}$ 
\citep{itoh2018bros} 
while the criterion was $\alpha>-0.5$ 
at the observation times in this paper. 
We carry out optical and NIR imaging observations to detect any brightness change 
possibly from a neutrino-emitting blazar within an IceCube localization. 
For this purpose, 1-2~m class telescopes in the OISTER collaboration 
are mainly used. 
A fraction of the BROS sources are apparently faint in optical and not detected in the PS1 data. 
We search for variabilities of such faint blazar candidates with Hyper Suprime-Cam (HSC; \cite{miyazaki2012}) 
on the 8.2-m Subaru telescope. 

A supernova is also thought to be 
one of the candidates 
to generate high-energy neutrinos. 
To find a supernova which neutrino may originate from, 
we also carry out wide-field imaging surveys in optical to 
cover a significant fraction of IceCube localization 
using wide-field optical imagers 
such as Kiso Wide Field Camera (KWFC; \cite{sako2012})\footnote{KWFC was decommissioned in August 2018. } 
and \tomoe\ \citep{sako2016,sako2018} on the 1.05-m Kiso Schimidt telescope 
and HSC on the 8.2-m Subaru telescope. 

For the \thisevent\ event, 
we performed optical and NIR observations 
with both the strategies. 
Our blind survey for supernovae is described in a separate paper 
(\morokumasnpaper).

\subsection{Optical and NIR Imaging}\label{sec:sec_optimg}

All the imaging observations are summarized below and in Table~\ref{tab:tab_obsimg}. 
Detailed characteristics of the telescopes and instruments are summarized in the table. 

\subsubsection{Initial Response to the Alert: Search for a rapidly variable blazar}\label{sec:sec_searchblazars}
We started follow-up imaging observations 
$0.8$~days after the alert of IceCube-170922A 
Hiroshima Optical and Near-Infrared camera (HONIR; \cite{akitaya2014})
on the 1.5-m Kanata telescope at the Higashi Hiroshima Observatory ($R_C$ and $J$-bands) and 
KWFC on the 1.05-m Kiso Schmidt telescope ($g,r$ and $i$-bands). 
In subsequent nights, 
HSC on the 8.2-m Subaru telescope ($z$-band) 
and 0.5-m MITSuME Akeno telescope 
(\cite{yatsu2007}; \cite{shimokawabe2008}; \cite{kotani2005}; 
$g, R_C$, and $I_C$-bands) were also used. 
Field-of-views of 
KWFC 
and HSC 
are well suited for effectively covering the localization area given by IceCube as shown in Figure~\ref{fig:fig_imgfield}. 

Brightness of a counterpart was unknown and difficult to predict 
because of unknown nature of a counterpart and unknown distance to a counterpart. 
Then, we adopt multiple exposure times to be 
3-180 seconds as summarized in Table~\ref{tab:tab_obsimg}. 

There were 7 flat-spectrum radio sources 
in a preliminary version of the BROS catalog 
within or right outside the 90\% localization of \thisevent\ 
as shown in Figure~\ref{fig:fig_imgfield} and 
as listed in Table~\ref{tab:tab_brossources}. 
Five sources among the seven are detected 
in the optical archival PS1 DR1 data. 
Each of the 7 BROS sources was observed basically one by one 
with Kanata/HONIR and MITSuME while 
all the seven BROS sources were covered by two KWFC pointings 
and two HSC pointings. 
As described in \S\ref{sec:sec_discnirvari}, 
we made a quick data reduction for the data 
and a Kanata/HONIR difference image for \target\ 
revealed that it was fading in a time scale of a day. 

\subsubsection{Monitoring for \target}\label{sec:sec_monitoring}
After detecting a rapid NIR variability of \target\ with Kanata/HONIR (described in \S~\ref{sec:sec_discnirvari}), 
we continued monitoring \target\ with telescopes described 
in \S\ref{sec:sec_searchblazars}. 
In addition, we carried out 
$J$, $H$, and $K_s$-band simultaneous imaging with 
SIRIUS \citep{nagayama2003,nagayama2012} on the 1.4-m Infrared Survey Facility (IRSF) and 
$V$-band imaging with the 0.4-m Kyoto telescope. 
Exposure times of the single images are 10 and 60~seconds, respectively. 
We also took non-filter CMOS imaging data 
with \tomoe\ \citep{sako2016,sako2018} on the Kiso Schmidt telescope. 
Two~fps (consecutive 0.5-sec exposure with negligible readout time) full-frame readout mode 
\citep{sako2018} was adopted. 
No filters were used 
and 
the sensors of \tomoe\ are sensitive 
in $\sim350-900$~nm \citep{kojima2018}. 

\begin{table*}[!htbp]
  \tbl{Telescopes and Instruments Used for Follow-up Imaging Observations.}{
  \begin{tabular}{cllllll}
      \hline
      Mode & Telescope & Aperture [m] & Instrument & FoV & Filter & $t_{\rm{exp}}$ [sec]\\ 
      \hline
      M  & Kyoto           & 0.4 & - & 18~arcmin (rectangle)  & $V$         & 60 \\
      SM & MITSuME (Akeno) & 0.5 & - & 28~arcmin (rectangle)  & $g,R_C,I_C$ & 60 \\
      SM & Kiso   & 1.05   & KWFC    & 2.2~deg (rectangle)    & $g,r,i$ & 10,30,60,180 \\
      M  & Kiso   & 1.05   & Tomo-e Gozen  & ($39.7\times22.4$~arcmin$^2$)$\times$4$^{*}$ & No      & 0.5 \\
      M  & IRSF   & 1.4    & SIRIUS  & 7.7~arcmin (rectangle) & $J,H,K_s$ & 10 \\
      SM & Kanata & 1.5    & HONIR   & 6~arcmin (diameter)    & $R_C,J$ & 25-95 ($R_C$), 10-80 ($J$) \\
      SM & Subaru & 8.2    & HSC     & 1.5~deg (diameter)     & $z$     & 3-5\\
	\hline
  \end{tabular}}
  \label{tab:tab_obsimg}
\begin{tabnote}
In the first column, S and M denote 
``survey'' and ``monitoring'', respectively. 
(*) indicates that 
when we took \tomoe\ data, 
the instrument 
was operated with the limited number (4) of the sensors 
before the completion of the instrument with 84~sensors in 2019. 
\end{tabnote}
\end{table*}

\begin{table*}[!htbp]
  \tbl{Seven flat-spectrum radio sources observed within or right outside the IceCube 90\%\ error region.
  }{%
  \begin{tabular}{lllllrrrrl}
      \hline
           &    RA  & Dec    & RA    & Dec   & sep.     & $f_{\rm{TGSS}}$ & $f_{\rm{NVSS}}$ &          & \\
      Name & (NVSS) & (NVSS) & (PS1) & (PS1) & [arcmin] & [mJy]           & [mJy]           & $r(PS1)$ & note\\
      \hline
      BROS~J0509+0541 &  05:09:26.0 & +05:41:35.6 & 05:09:26.0 & +05:41:35.4 &   4.58 &  406.7 &  546.8 & 15.04 & TXS~0506+056, FL8YJ0509.4+0542\\
J0509+0529      &  05:09:12.0 & +05:29:22.0 & 05:09:12.3 & +05:29:22.8 &   15.83 &   25.8 &   17.9 & 
17.05 & No objects in latest BROS\\ 
BROS~J0512+0538 &  05:12:05.7 & +05:38:41.7 & 05:12:05.7 & +05:38:41.2 &  35.72 &   57.9 &   24.0 & 15.95 & None\\
BROS~J0512+0540 &  05:12:11.6 & +05:40:15.2 & 05:12:11.6 & +05:40:15.6 &  37.03 &  160.9 &   63.0 & 22.62 & None\\
J0512+0615      &  05:12:36.0 & +06:15:45.0 & 05:12:36.6 & +06:15:43.8 &  54.03 &   24.7 &   10.2 & 
20.86 & No objects in latest BROS\\ 
BROS~J0512+0608 &  05:12:56.7 & +06:08:19.1 & 05:12:56.8 & +06:08:17.6 &  54.28 &  350.9 &  159.5 & 18.46 & None\\
BROS~J0514+0549 &  05:14:40.1 & +05:49:23.2 & 05:14:39.5 & +05:49:20.3 &  74.11 &  267.6 &  268.0 & 19.87 & None\\
	\hline
    \end{tabular}}\label{tab:tab_brossources}
\begin{tabnote}
  These are registered in the preliminary version of the BROS catalog when we started the follow-up observations. 
  Coordinates of the sources in the NVSS and PS1 catalogs, 
  separation from the IceCube detection center in arcmin unit, 
  radio fluxes in 150~MHz (TGSS) and 1.4~GHz (NVSS) in mJy unit, 
  PS1 $r$-band Kron magnitude, and some notes are listed. 
\end{tabnote}
\end{table*}

\begin{table*}[!htbp]
  \tbl{Telescopes and Instruments Used for Follow-up Spectroscopic Observations.}{
  \begin{tabular}{llllllrlll}
      \hline
      Telescope & Aperture [m] & Instrument & Grism & Filter & $\lambda$~[\AA] & $R$ & $t_{\rm{exp}}$ [sec] & Date (UT) & MJD \\
      \hline
      Nayuta       & 2.0 & MALLS  & 150~l/mm & WG320 & 4700--8600  & $600$ & $900\times6$ & 2017-09-29 & 58025.7 \\
      Kanata       & 1.5 & HOWPol & grism    & NONE  & 4700--9200  & $400$ & $900\times3$ & 2017-09-29 & 58025.7 \\
      Subaru       & 8.2 & FOCAS  & 300B     & SY47  & 4600--8200  & $400$ & $100\times11$& 2017-09-30 & 58026.6 \\
      Subaru       & 8.2 & FOCAS  & VPH850   & SO58  & 7500--10500 & $1200$& $100\times7$ & 2017-10-01 & 58027.6 \\
      Gemini-North & 8.2 & GMOS   & B1200    & NONE  & 3400--4960  & $3700$& $600\times(3+8)+240\times1$ & 2017-11-11,15 & 58071.5\\
	\hline
    \end{tabular}}\label{tab:tab_obsspec}
\begin{tabnote}
$\lambda$ indicates usable wavelength ranges rather than observed wavelength ranges. 
$R$ indicates spectral resolutions. 
\end{tabnote}
\end{table*}

\subsubsection{Data Reduction of Imaging Data}
\label{sec:sec_dataredimg}
All the data are reduced in a standard manner, 
including 
bias, overscan, and dark current subtractions if necessary and possible, 
flat-fielding, 
astrometry with 
Optimistic Pattern Matching (OPM; \cite{tabur2007}) 
or 
Astrometry.net \citep{lang2010} 
although some differences exist; 
e.g., distortion correction is applied only for HSC data. 
Details of the data reductions are described 
in \citet{morokuma2014} for KWFC, 
in 
\citet{tachibana2018} for MITSuME. 
The HSC data are analyzed with {\it hscPipe} v4.0.5, which is a standard reduction pipeline 
of HSC \citep{bosch2017}. 
\tomoe\ CMOS sensor data are also reduced with a dedicated pipeline described in \citet{ohsawa2016}. 

To see intranight variability of objects, 
we measured apparent magnitudes of \target\ and neighboring stars in individual frames. 
With these photometry results, we obtain daily-averaged magnitudes to obtain daily light curves of \target\ and neighboring stars. 

Photometry was done with SExtractor \citep{bertin1996} and MAG\_AUTO is used. 
All magnitudes in optical and NIR wavelengths are measured in AB system. 
Optical magnitudes are calibrated 
relative to the PS1 Data Release~2 (DR2) catalog (\cite{magnier2016}; \cite{flewelling2016}). 
Johnson-Cousins filter data are also calibrated to PS1 data in filters 
whose bandpasses are similar 
(i.e., PS1~$r$ for $R_C$ and PS1~$i$ for $I_C$). 
No filter \tomoe\ data are also calibrated relative to $r$-band PS1 data. 
Magnitudes of field stars in NIR wavelengths 
are derived from the 2MASS database~\citep{skrutskie2006} 
and converted to those in the AB system by following \citet{tokunaga2005}. 

To 
search for a counterpart (\S\ref{sec:sec_searchblazars}), 
we apply an image subtraction method (\cite{alard1998}; \cite{alard2000}) 
for the imaging data 
with reference images of PS1 in optical and 2MASS in NIR, respectively. 
Except for Subaru/HSC, all of our optical imaging data are shallower than PS1 images and 
the depths of the search are limited by the depths of our data. 
On the other hand, 
in NIR, 2MASS data are not so deep and 
we also performed another subtraction in which 
our first (reference) images are subtracted from our new data. 

For photometry for the normal images without image subtractions, 
we add addtional 3\% errors to the errors measured with SExtractor. 
This is because photometry errors based on photon statistics usually 
underestimate the errors which should follow Gaussian distribution around the real values for non-variable objects, 
for example, due to imperfect flat-fielding procedure. 

\begin{figure}
 \begin{center}
  \includegraphics[width=96mm,bb=0 0 466 792]{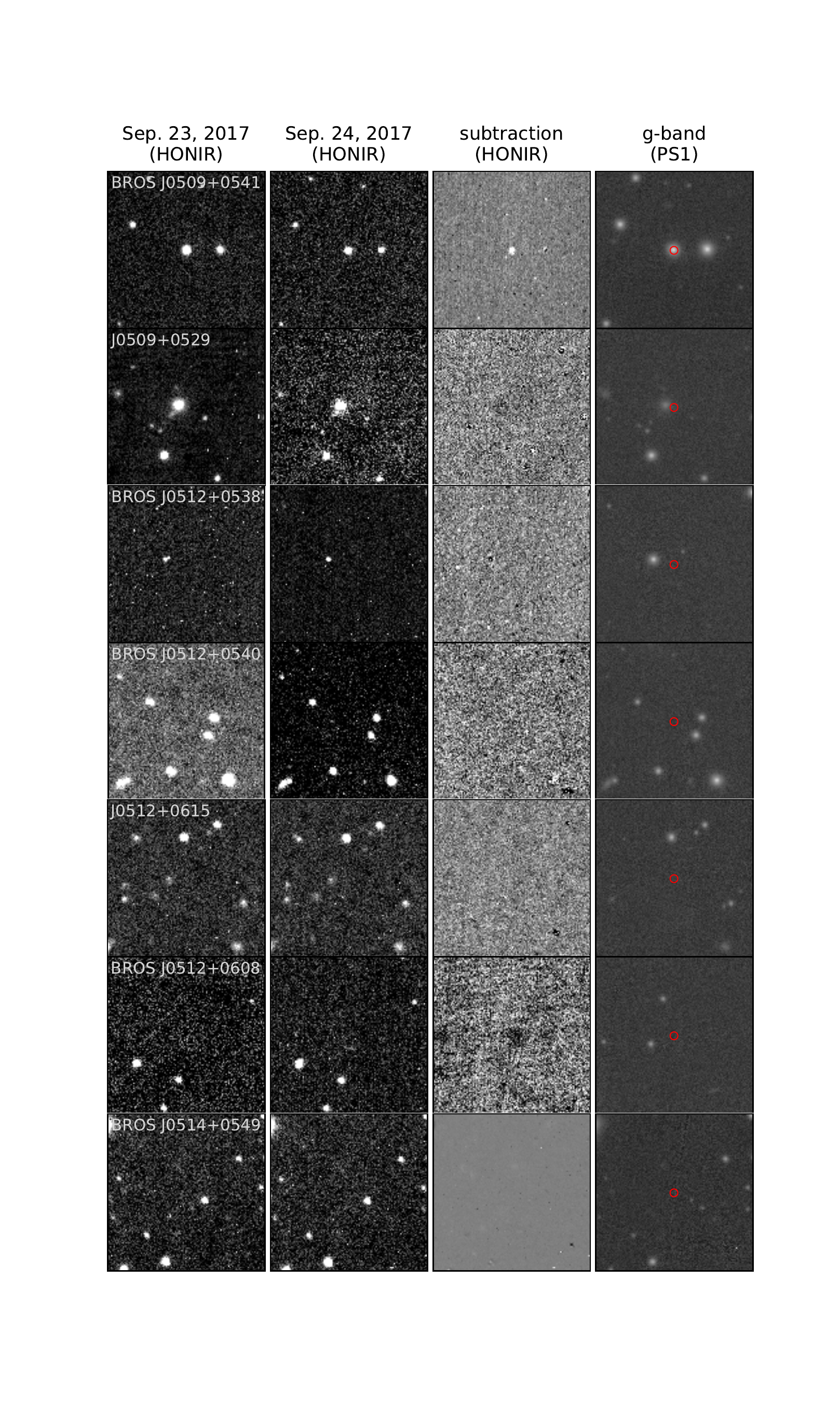}
 \end{center}
\caption{
HONIR $J$-band images and subtracted images for the seven 
flat-spectrum radio sources catalogued in the preliminary BROS 
when we started the follow-up observations 
(Table~\ref{tab:tab_brossources}). 
From left to right, 
HONIR $J$-band image on Sep. 23, 2017, 
HONIR $J$-band image on Sep. 24, 2017, 
subtracted HONIR $J$-band image (Sep. 23 - Sep. 24), 
and 
PS1 $r$-band image are shown for each BROS source. 
Red circles on the PS1 images are NVSS radio locations with radii of 2~arcsec, 
which are typical positional errors of NVSS sources. 
}\label{fig:fig_discimg_bros}
\end{figure}

\begin{figure*}
 \begin{center}
	\includegraphics[width=165mm,bb=0 0 654 282]{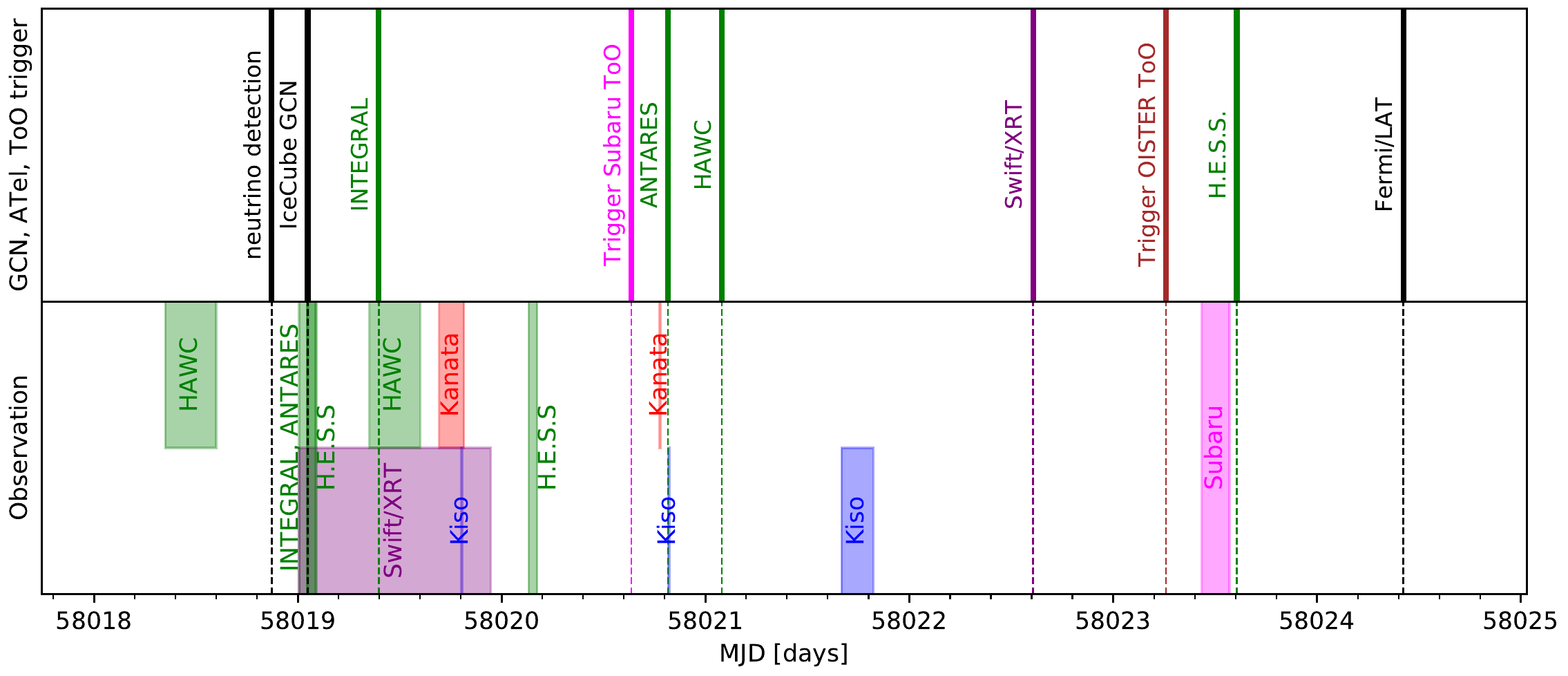}
 \end{center}
\caption{
Timeline of the follow-up observations from the IceCube alert. 
Kanata/HONIR, Kiso/KWFC, and Subaru/HSC observations 
were shown in red, blue, and magenta, respectively. 
The IceCube neutrino detection, its alert, and the first Fermi/LAT GCN 
indicating \target\ is a possible counterpart 
by \citep{tanaka2017_IC170922a} are indicated 
in solid black at the top panel. 
}\label{fig:fig_timeline}
\end{figure*}

\subsubsection{ASAS-SN monitoring data}
\label{sec:sec_asassndata}

ASAS-SN is covering a large fraction of the sky.
Long-term data nicely covering before and after the neutrino detection 
are available and significant variablity and brightening of \target\ 
before the neutrino detection were reported in \citet{franckowiak2017}. 
Original ASAS-SN $V$-band data 
are available and taken from 
ASAS-SN Sky Patrol\footnote{https://asas-sn.osu.edu} 
(\cite{shappee2014}; \cite{kochanek2017}). 
The data are 
calibrated relative to 
AAVSO Photometric All-Sky Survey (APASS; \cite{henden2014}). 
As mentioned in \citet{icecube2018_science1}, 
the nearby west star\footnote{separated by 17.079~arcsec, 
objID=114830773534352391, (RA, Dec) = (05:09:24.82, +05:41:35.7) in PS1 DR2.} 
($g_{\rm{MeanPSF}}=14.7782\pm0.0024$, $r_{\rm{MeanPSF}}=14.4373\pm0.0010$, 
in the PS1 DR2 catalog, corresponding to $V=14.587$, which is converted using an equation 
in \cite{kostov2018}) flux 
contaminates into the target flux 
in photometry of 
ASAS-SN Sky Patrol 
because of the coarse spatial sampling (8~arcsec pixel$^{-1}$) 
and large PSF ($\sim15$~arcsec FWHM) 
of ASAS-SN. 
We estimate the contribution to the \target\ to be 30\% of the nearby star flux 
by calculating the contaminating flux of the nearby star 
into the aperture of \target\ 
with the PSF size, shape (circular), and aperture adopted for the ASAS-SN photometry 
and 
subtract it 
from the target flux 
to extract only the flux of \target. 
We here do not add additional errors to the original error of \target. 
We also confirm this correction does not change our conclusions. 
The brightness of the nearby west star is 
confirmed to be almost constant over our observing period based on our data.

\subsection{Optical Spectroscopy}\label{sec:sec_optspec}

An obstacle to study the neutrino source in detail was uncertain 
redshift determination of \target. 
Redshift determination of BL~Lac objects is in general difficult 
and require a large amount of observing time with a large aperture optical telescope 
(\cite{landoni2015a}; \cite{landoni2015b}) 
even if their images can be easily taken with 1m-class telescopes. 
This was the case for \target\ and the redshift had not been reliably determined \citep{halpern2003}. 

\target\ was spectroscopically observed in the optical wavelength 
and the redshift is 
reported to be $z=0.336$ in \citet{ajello2014}. 
However, the origin of this redshift determination is unclear from the figure and text in the paper. 
MAGIC high-energy $\gamma$-ray detection (\cite{mirzoyan2017}; \cite{mirzoyan2018})
gives another constraint 
on the redshift 
through the measurement of $\gamma$-$\gamma$ 
attenuation effect. 
In \citet{icecube2018_science1}, several conservative estimates are done with the MAGIC data and 
$95$\%~confidence level upper limit on the source redshift is $z<0.98$ 
while the lowest upper-limit redshift is $z<0.41$. 
Since there were 
still debates on the redshift determination 
before \citet{paiano2018} determined the redsfhit to be $z=0.3365$, 
spectroscopic measurement was still important (\cite{steele2017}; \cite{morokuma2017_IC170922a}).

We took optical long-slit spectra with 
Medium And Low-dispersion Long-slit Spectrograph (MALLS) on the 2-m Nayuta telescope, 
HONIR \citep{akitaya2014} on the 1.5-m Kanata telescope, 
Faint Object Camera and Spectrograph (FOCAS; \cite{kashikawa2002}) on the 8.2-m Subaru telescope, 
and 
Gemini Multi-Object Spectrograph (GMOS; \cite{hook2004}) on the 8.2-m Gemini-North telescope. 
The spectral resolutions of the observations are mostly as low as $R\sim1000$ or even lower. 
The exposure times are not so long, 1.5~hours at longest. 
These are summarized in Table~\ref{tab:tab_obsspec}. 
The FOCAS observations and spectra are described and shown in the supplement part of \citet{icecube2018_science1}. 

All the spectra are reduced in a standard manner with IRAF, 
including bias subtraction, flat-fielding, wavelength calibration, 
sky subtraction, and source extraction, 
while the Gemini-N/GMOS data is reduced with the Gemini IRAF package. 
Wavelength calibration was done using lamp spectra or night sky lines. 
Flux calibration is not applied and the obtained 1d spectra are normalized with continua 
to see any weak emission and absorption lines. 

\subsection{Optical and NIR Polarimetry}\label{sec:sec_optnirpol}

We also conducted optical and NIR polarimetry observations with HONIR \citep{akitaya2014} 
at the Cassegrain focus on the 1.5-m Kanata telescope. 
HONIR is equipped with a rotatable half-wave plate and a Wollaston prism 
which enables us to conduct 
simultaneous polarimetry observations 
in an optical and an NIR channels. 
We used  
$R_C$-band and $J$-band in the optical and NIR channels, respectively. 

The data were taken on 15~nights, 
from $t=8$~days to 
$t=213$~days 
after the IceCube alert ($t=0$). 
Each observation consisted of a set of 4~exposures at half-wave plate position angles 
of 
$0^{\circ}.0$, 
$22^{\circ}.5$, 
$45^{\circ}.0$, and 
$67^{\circ}.5$. 

The data were reduced 
following a methodology of data analysis described in \citet{kawabata1999} 
to derive polarization degrees and angles. 
Instrumental polarization induced by the optical system 
within the Kanata telescope and HONIR has been confirmed to be 
as small as 0.1-0.2\% \citep{akitaya2014,itoh2017}. 

Data taken with a fully-polarizing filter inserted in front of HONIR 
are used for the correction of the wavelength-dependent origin points 
of the position angles (originated from the multi-layered superachromatic
half-wave plate). 
The obtained polarization degrees are 
as stably high as $\gtrsim99$\%\ 
with this filter in both $R_C$ and $J$ bands 
and we do not perform the depolarization correction.
A strongly polarized star, BD+64d106 \citep{schmidt1992}, 
was observed to correct
observed position angle in the celestial coordinate 
and is used to calibrate the position angles of \target. 
Galactic foreground polarization should be almost negligible 
($P_{R}\lesssim 0.7$\%\ or $P_{J}\lesssim0.2$\%) 
because the interstellar extinction toward \target\ 
is $A_{R}=0.235$ or $A_{R}=0.077$~mag \citep{schlafly2011,serkowski1975}. 
These values are mostly smaller than the measured values 
and then we do not adopt any correction for the observed polarization. 

Observing epochs are separated into two as below. 
The first epochs are $\sim1.5$~months after the IceCube alert. 
The second epochs are around 
$t\sim180$~days after the alert, 
when larger polarization degrees of \target\ 
based on polarimetric data taken with Liverpool/RINGO 
are reported \citep{steele2018}. 
Motivated by this report, we also 
took additional polarization data with Kanata/HONIR. 

\section{Results and Discussions}\label{sec:sec_resultsdiscussion}

\subsection{Discovery of Rapid NIR Variability of \target}\label{sec:sec_discnirvari}

We examined the subtracted $J$-band HONIR images 
(HONIR-HONIR and HONIR-2MASS) to see any rapid variability of 
the BROS blazars and blazar candidates. 
Figure~\ref{fig:fig_discimg_bros} shows 
HONIR-HONIR subtraction images 
for the 7 sources 
in Table~\ref{tab:tab_brossources}, which were listed 
in the preliminary BROS catalog at the observation time. 
We found that \target\ showed a fading trend, by $\sim0.15$~mag, 
from Sep. 23 to Sep. 24 in 2017 
as shown in the top panels of Figure~\ref{fig:fig_discimg_bros}. 
In $g$-band data taken with Kiso/KWFC, about $0.15$~mag decline was also detected. 
For the other sources, we did not find any significant NIR variability 
from the 2-night HONIR data 
or did not detect the object, which was consistent with 
non-detection or faint magnitudes recorded in the PS1 optical imaging data. 

This rapid brightness change of \target\ 
may indicate a possible relation
with the neutrino detection, 
motivating examining {\it Fermi} $\gamma$-ray all-sky monitoring data. 
One of the co-authors of this paper led this effort, 
found its $\gamma$-ray variability, 
and reported it 
via The Astronomer's Telegram (ATel) \citep{tanaka2017_IC170922a}. 
This was further followed by 
multi-wavelength follow-up observations 
all over the wavelengths (Figure~\ref{fig:fig_timeline}; \cite{icecube2018_science1}). 

Timeline from the IceCube alert \citep{kopper2017icecubegcn} 
to the first Fermi ATel report \citep{tanaka2017_IC170922a} is summarized in 
Figure~\ref{fig:fig_timeline}. 
After the IceCube alert, 
some observational reports 
with monitoring and follow-up data 
are distributed via GCN and ATel. 
In summary, 
no possible related objects to the IceCube neutrino were mentioned in any of the reports before \citet{tanaka2017_IC170922a}. 
At the event time, 
no significant $\gamma$-ray (INTEGRAL SPI-ACS in \cite{savchenko2017}; HAWC in \cite{martinez2017}; H.E.S.S. in \cite{denaurois2017atel}), 
no significant neutrino (ANTARES in \cite{dornic2017,dornic2017atel}), 
were detected. 
In Swift/XRT follow-up observations, 
nine sources were detected (including 8 known sources) \citep{keivani2017} 
although no special notices were made for \target. 

Optical long-term data in this field had been taken in the 
ASAS-SN 
project since October, 2012. 
After the report by \citet{tanaka2017_IC170922a}, 
\target\ was reported by  
\citep{franckowiak2017}
to be at the brightest phase in the ASAS-SN data 
and to show a brightening of $\sim0.5$~mag in $V$-band over the last 50 days. 

\subsection{Light Curves of \target}\label{sec:sec_lc}

\begin{figure*}
 \begin{center}
  \includegraphics[width=116mm,bb=0 0 566 1077]{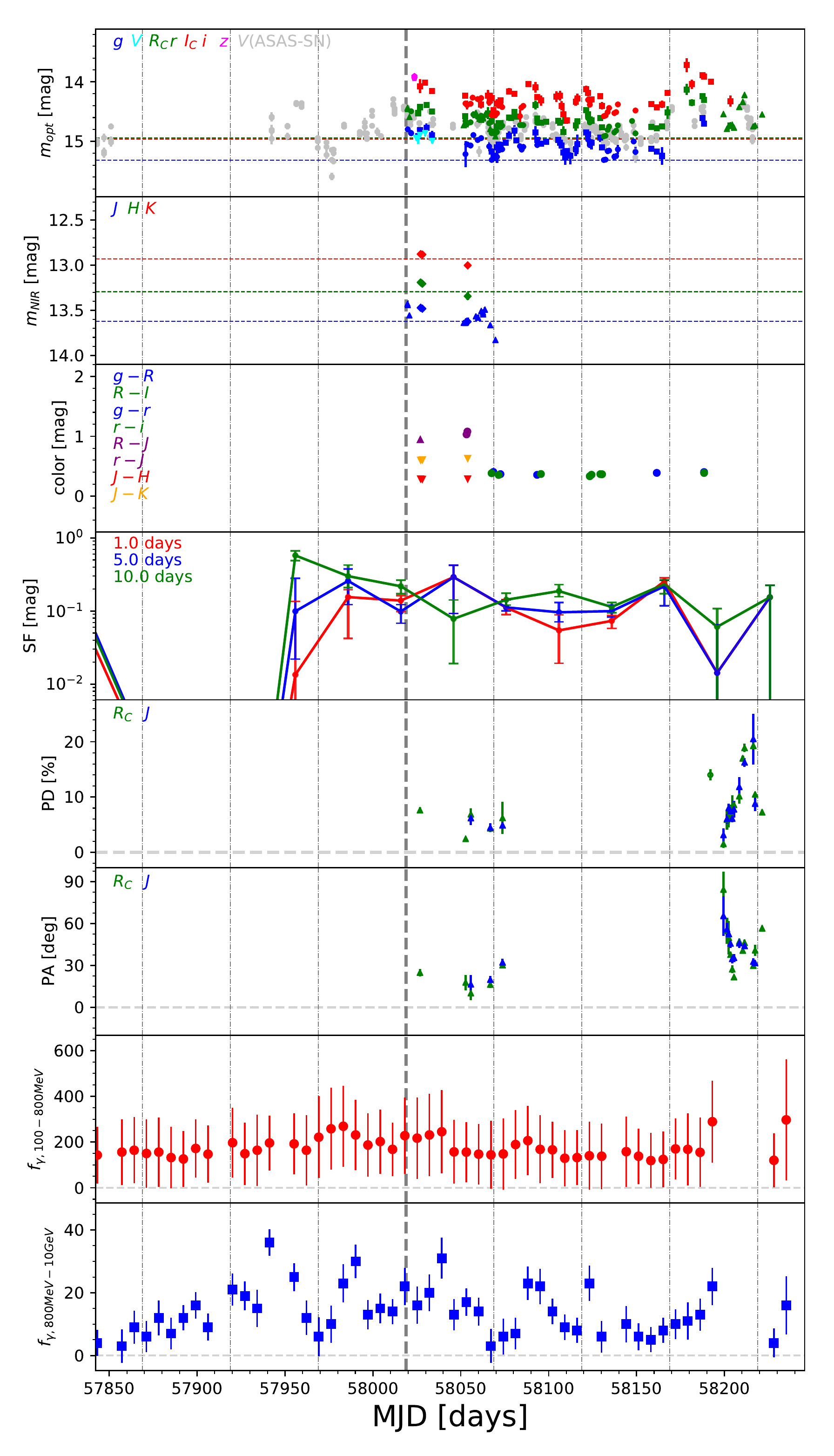}
 \end{center}
\caption{
Daily variabilities of 
optical ($g$; blue, $V$: cyan, $V$ by ASAS-SN: gray, 
$R_C$ and $r$: green, $I_C$ and $i$: red, $z$: magenta)
and NIR ($J$: blue, $H$: green, $K_s$: red) fluxes, 
optical-NIR colors 
from the top to the 3rd panel. 
Changes of 
the structure functions in time scales of 1.0, 5.0, and 10.0~days 
are shown in the 4th panel. 
Daily changes of polarization degree ($R_C$: green, $J$: blue), 
polarization angle ($R_C$: green, $J$: blue), 
Fermi/LAT $\gamma$-ray fluxes in 200-800~MeV and 800~MeV-10~GeV energy bands 
are shown from the 5th to the bottom panel. 
In the polarization degree, Liverpool/RINGO3 data is plotted in open circles (MJD$\sim58192$). 
The neutrino detection time is indicated as gray dashed vertical lines. 
Vertical dash-dot gray lines indicate $-150, -100, -50, +50, +100, +150, +200$~days 
with respect to the IceCube neutrino detection. 
Galactic extinction is not corrected. 
}\label{fig:fig_lc}
\end{figure*}

\begin{figure*}
 \begin{center}
  \includegraphics[height=207mm,bb=0 0 407 532]{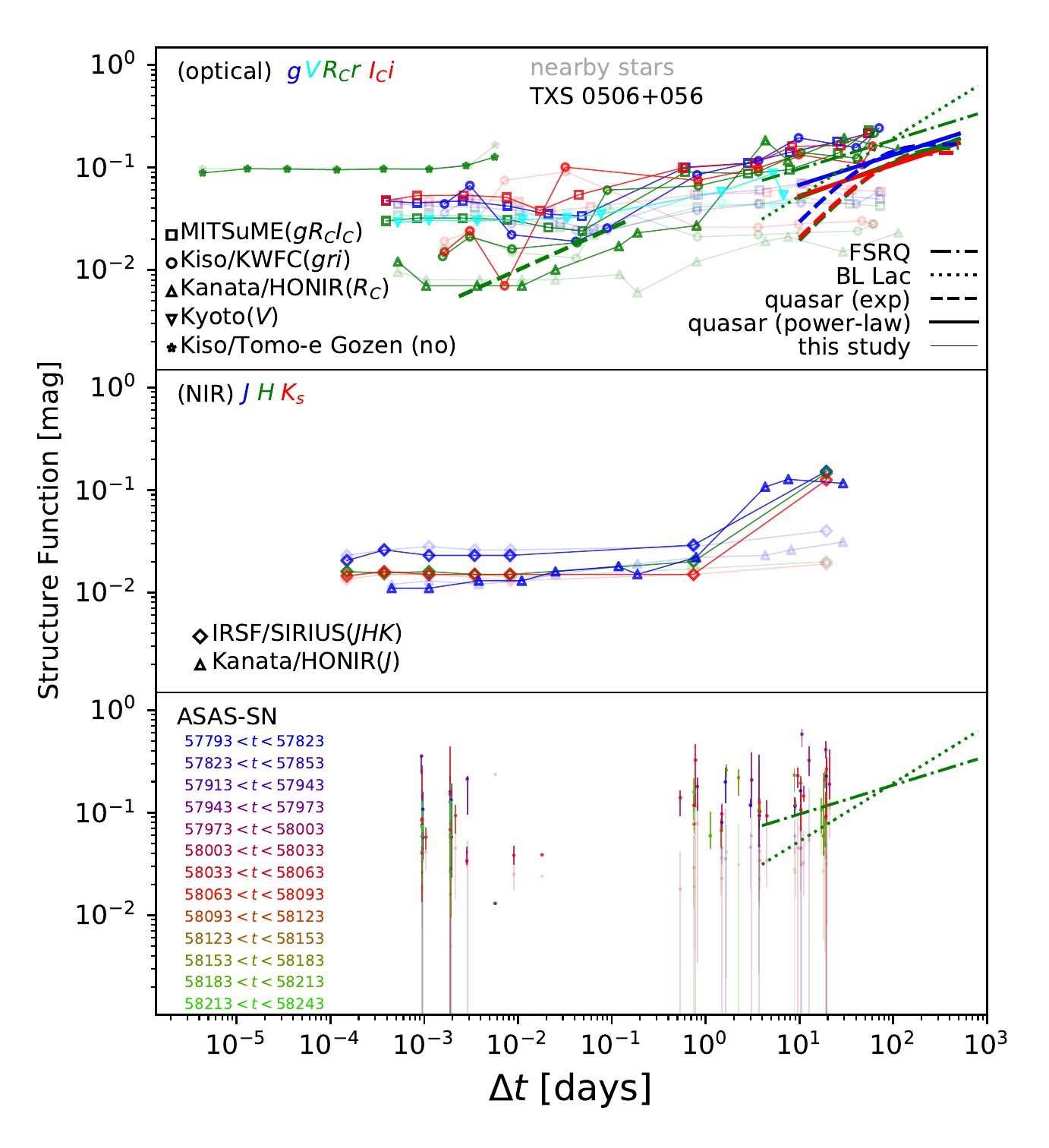}
 \end{center}
\caption{
Optical (top) and NIR (middle) structure functions of \target\ 
obtained in our observations in thick points and lines. 
SFs of the neighboring stars are also shown in thin points and lines 
to see how large the SF measurement errors are. 
obtained in our observations in thick points and lines. 
Power-law and exponential fitting results 
for quasars \citep{vandenberk2004} are shown 
in blue ($g$), green ($r$), and red ($i$) thick solid and dashed lines in $\Delta{t}>10^{1}$~days, respectively. 
Power-law fitting results for FSRQs and 
BL~Lacs \citep{bauer2009blazar} are also shown 
in green dotted-dashed and dotted lines, respectively. 
Flat parts in short time scales are dominated by measurement errors and 
real SF amplitudes are thought to be lower than these lines. 
Optical ASAS-SN structure function as a function of period (every 30~days) 
is also shown in the bottom panel. 
}\label{fig:fig_sf}
\end{figure*}

Time variabilities of the observables are shown in Figure~\ref{fig:fig_lc}, 
including those of 
optical magnitudes, 
NIR magnitudes, 
optical and NIR colors, 
optical and NIR polarization degrees and angles, and 
$\gamma$-ray fluxes in the low (100-800~MeV) and high (800~MeV-10~GeV) energy bands. 
In addition to our own data (\S\ref{sec:sec_optimg}), 
optical $V$-band data from 
ASAS-SN (see \S\ref{sec:sec_asassndata}), 
RINGO3 optical polarimetry data (see \ref{sec:sec_optnirpol}), 
and {\it Fermi} $\gamma$-ray data are also used. 

The $\gamma$-ray fluxes are taken from the Fermi All-sky Variability 
Analysis (FAVA; \cite{abdollahi2017})\footnote{https://fermi.gsfc.nasa.gov/ssc/data/access/lat/FAVA/} website. 
Aperture photometry fluxes in the low (100-800~MeV) and high (800~MeV-30~GeV) energy bands 
are used in this paper. 
We understand that the FAVA light curves are preliminary as described on the FAVA website but 
the temporal behavior is similar to those shown in \citet{icecube2018_science1} and 
the FAVA data are 
good enough for our purpose. 

\subsection{Variability of \target}\label{sec:sec_vari}

We investigate 
optical and NIR 
variabilities in three time scales 
including 
daily (\S\ref{sec:sec_longtermvari}), 
intranight (\S\ref{sec:sec_intranightvari}), 
and second-scale (\S\ref{sec:sec_secondvari}) ones. 
We construct 
structure functions (SFs) 
of optical and NIR variability of \target, 
which is, in general, defined to be an ensemble variability of an object or a specific set of objects, 
to quantify its time variability. 
We compare SFs of \target\ with 
those of other AGN in the literature. 
A caveat and problem are pointed out in \citet{emmanoulopoulos2010} when an SF is used 
for studying a time scale of blazar variability, 
however, in this paper, we discuss only the variability amplitudes at given time scales, 
not the time scale itself. 

We adopt an usual concept on 
a definition of the structure function (SF) $V(\Delta{t})$ as below, 
\begin{eqnarray}
	V(\Delta{t})&=&\sqrt{<|\Delta{m}|^2>-<\sigma^2_{S/N}>}
\end{eqnarray}
where $\Delta{t}$ is a time interval between observations, and 
$\Delta{m}$ is a magnitude difference between different observations 
and $\sigma_{\rm{S/N}}$ is measurement errors in magnitude unit. 
We calculate the SFs 
for \target\ and neighboring stars with similar brightness 
to \target.
These two SFs 
are compared to see any significant flux variability of \target. 
We estimate median and $1\sigma$ confidence level 
with a bootstrap method. 

The ASAS-SN data are useful for evaluating 
variability behavior before and around the alert while 
our own data enable us to investigate short time scale variabilities after the alert. 

Obtained SFs 
of \target\ are shown in thick points and lines 
in Figure~\ref{fig:fig_sf}. 
SFs of the neighboring stars are also shown 
in thin faint points and lines in Figure~\ref{fig:fig_sf}. 
The SFs of \target are 
compared with those of 
blazars (BL~Lacs and FSRQs) studied in the Palomar-QUEST survey \citep{bauer2009blazar}, 
SDSS quasars \citep{vandenberk2004}, 
and CTA102 \citep{bachev2017}. 
In \citet{bauer2009blazar}, 
94 BL~Lacs, 278 FSRQs, and 4 marginally classified (BL~Lacs or FSRQs) objects in total are observed 
with 48-inch Samuel Oschin Schmidt Telescope and 
QUEST2 Large Area Camera ($4.6^{\circ}\times3.6^{\circ}$ field-of-view; \cite{baltay2007})
over 3.5~years to construct the rest-frame SFs 
of them as shown in Figure~\ref{fig:fig_sf}. 
In \citet{vandenberk2004}, $>25,000$ quasars are observed in the 5 optical bands ($ugriz$) 
and rest-frame (both in time scales and wavelengths) SFs 
are derived. 
These two SFs are derived with different formulations but quantitatively equivalent with each other. 

\subsubsection{Long-Term Variability}\label{sec:sec_longtermvari}
Daily or longer time scale variability is summarized in this subsection. 
First, optical and NIR fluxes are significantly variable as shown in Figure~\ref{fig:fig_lc}. 
The peak-to-peak amplitude reaches up to 1~mag. 
We also see some fluctuations in a time scale of a few days.
In addition, optical or NIR fluxes 
are not at their peaks at the neutrino detection as well as in $\gamma$-ray \citep{icecube2018_science1}. 
The overall trend around the neutrino detection indicates that \target\ is 
in a declining phase. 
About 180 days after the neutrino detection, 
optical and NIR brightness increase again up to a brighter level 
than that around the neutrino detection, 
however, no neutrino detection is not reported in this period. 

The optical SF amplitudes of \target\ 
are comparable with or slightly larger than 
those of AGN in the previous studies
in time scales of $\Delta{t}>10^{1}$~days. 
\target\ is more variable by a factor of $\sim2$ than the SDSS quasars 
on average \citep{vandenberk2004}. 
The SFs of \target\ in our measurements are larger than or comparable to 
those of FSRQs and BL~Lacs \citep{bauer2009blazar} at their face values. 
Note that the SFs of AGN in general have 0.1~dex or larger scatter (\cite{vandenberk2004}; \cite{bauer2009blazar}), 
which make the envelope of the distribution is overlapped 
between those of FSRQs and BL~Lacs 
and our measurements. 

For NIR variability, 
\target\ in $J$-band in time scales from a few days to a few tens days, 
and $H$ and $K_s$-band in a time scale of a few tens days 
are significantly variable than neighboring stars. 
The NIR variabilities of \target\ 
are 
as large as optical ones in these time scales.
NIR variabilities in shorter time scales than a few days 
are comparable to those of neighboring stars and are 
almost at the limit of the measurement errors.
Typical 
NIR variabilities of blazars in the literature are 
$\sim0.1$~mag (\cite{sandrinelli2014}; \cite{li2018}), 
which is comparable to our measurement for \target\ 
and this variability behavior of \target\ is not special. 

We also calculated SFs 
using the ASAS-SN $V$-band data
in each 30~days 
from MJD=57793 (225~days before the neutrino detection) 
to MJD=58243 (225~days after the neutrino detection), 
covering the IceCube neutrino detection time 
on MJD$\sim58018.87$.
Variability amplitudes in three time scales (1, 5, and 10~days) are 
derived from the SFs and 
shown in the 4th panel of Figure~\ref{fig:fig_lc}. 
As a whole, there are no special periods 
when significantly larger variability is detected than other periods. 
It is not 
clear but 
the SFs marginally 
indicates that 
long-term (10~days) variability around $\sim60$~days before the neutrino detection is the largest 
and larger than that in the neutrino detection period 
with significance of $2.1\sigma$. 
On the other hand, the short-term (1~day) variabilities are constant in time. 
This might indicate the neutrino emission would be related to the 10-days-timescale variability in this epoch.  
The hard $\gamma$-ray fluxes are also highly variable around this epoch (Figure~\ref{fig:fig_lc}). 

\begin{figure}[!htbp]
 \begin{center}
  \includegraphics[width=83.5mm,bb=0 0 396 432]{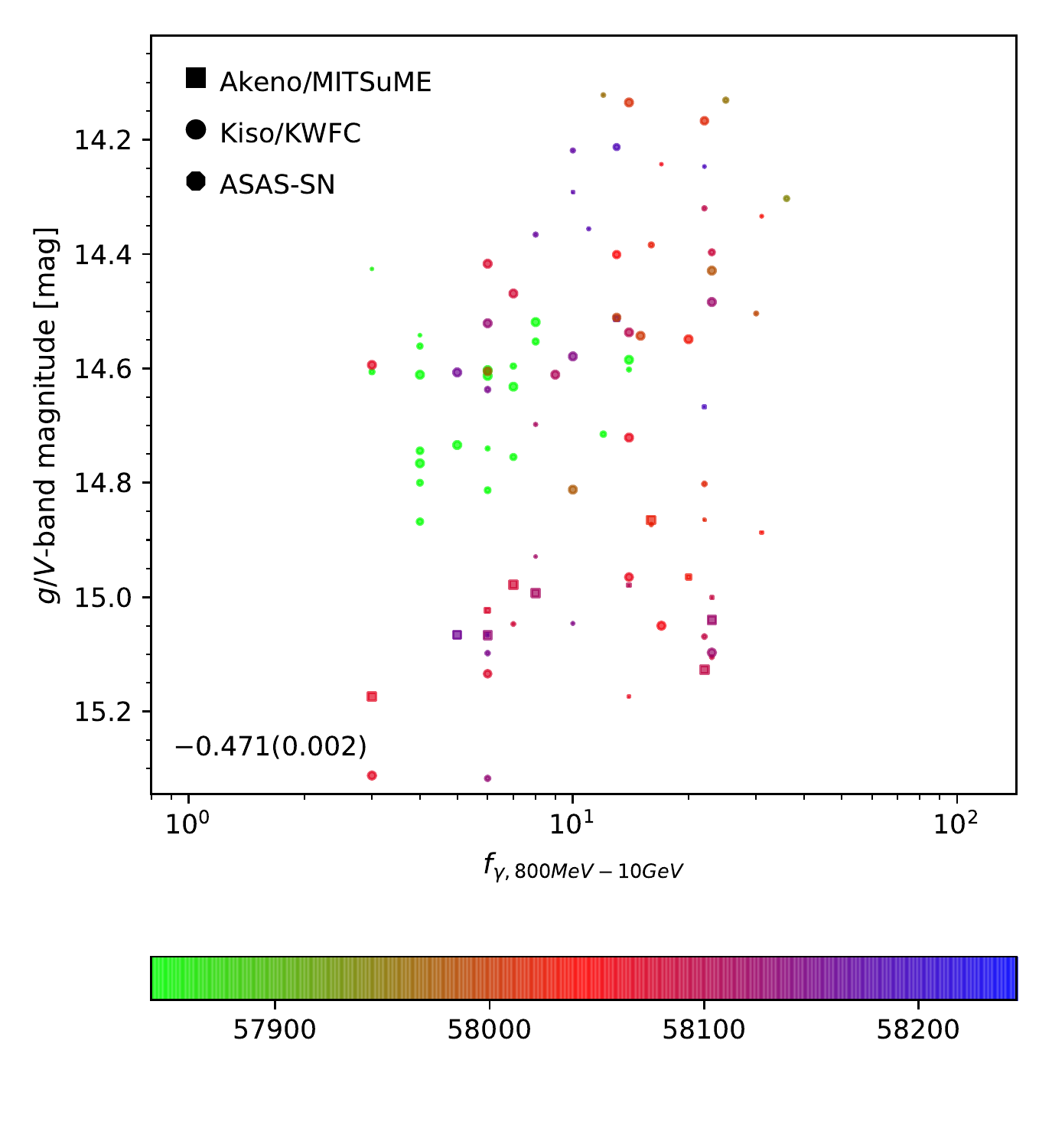}
 \end{center}
\caption{
$g$ or $V$-band magnitude 
as a function of 
$\gamma$-ray flux. 
Colors of the points indicate 
observing epochs (MJD) as shown in the color bar. 
}\label{fig:fig_optmag2bgamma}
\end{figure}

Correlation between 
the hard $\gamma$-ray fluxes (800~MeV-10~GeV) 
and optical brightness is investigated 
as seen in 
Figure~\ref{fig:fig_optmag2bgamma}. 
Optical magnitudes (brightness) are negatively (positively) 
correlated with the $\gamma$-ray fluxes with 
Spearman rank correlation coefficients of $-0.471$ and 
$p$-values of $<0.03$. 
This indicates that the correlation is significant and \target\ is brighter in optical 
in brighter $\gamma$-ray phases, which is consistent with general trend of ISPs 
or all types of blazars \citep{itoh2016,jermak2016}. 

\subsubsection{Intranight Variability}\label{sec:sec_intranightvari}
During 22 nights, 
we contiguously took imaging data of \target\ 
for a few tens minutes or longer 
with \kyototel, 
MITSuME, 
Kanata/HONIR, 
and IRSF/SIRIUS. 
With these datasets, intranight variability can be investigated. 

\begin{figure*}
 \begin{flushleft}
  \includegraphics[width=221.9mm,bb=0 0 814 539]{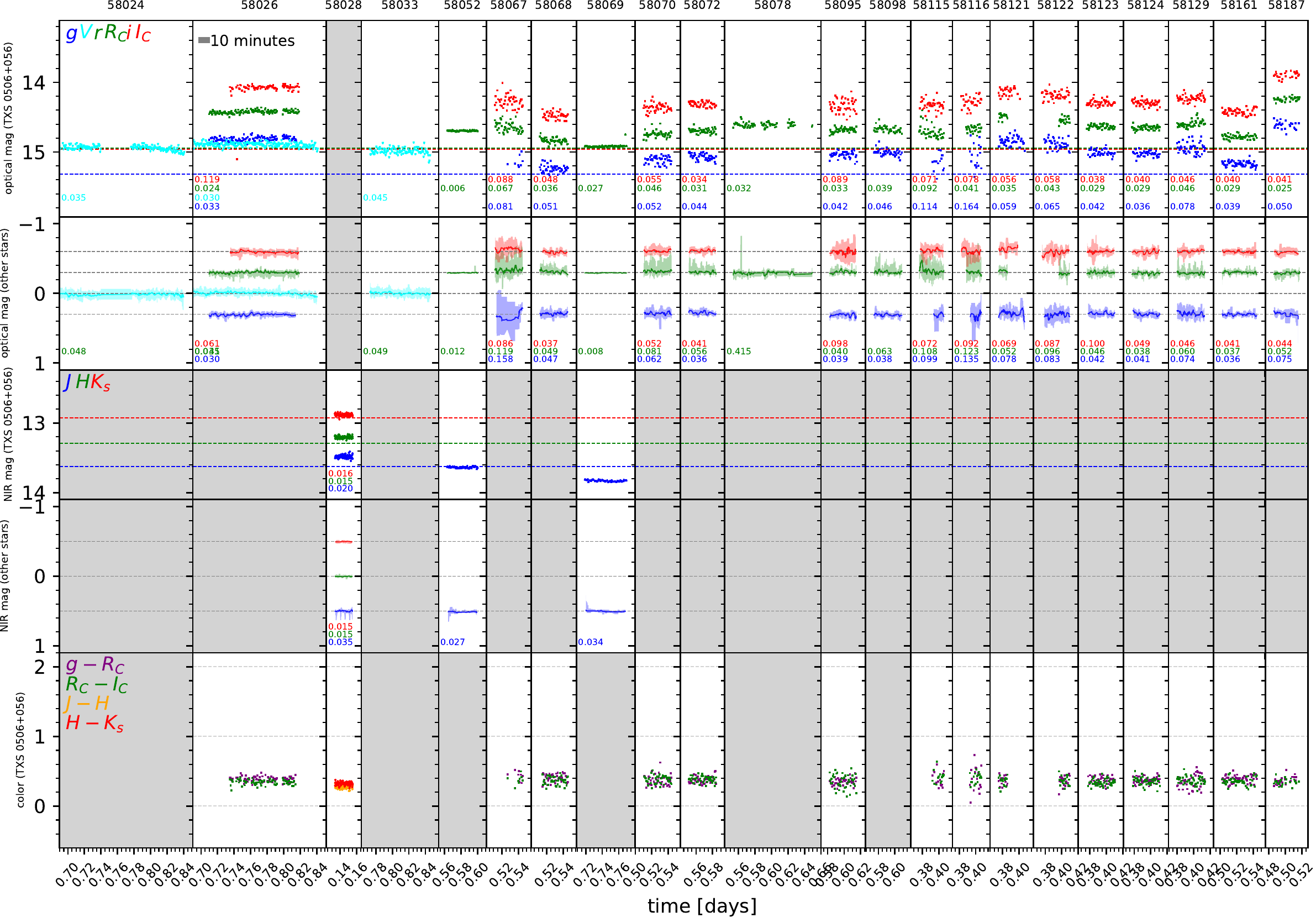}
 \end{flushleft}
\caption{
Light curves of \target\ based on photometry for each frame. 
From top to bottom, 
optical light curves of \target, 
optical light curves of nearby similarly bright stars, 
NIR light curves of \target, 
NIR light curves of nearby similarly bright stars, 
and optical and optical-NIR colors of \target\ are shown. 
Dates in MJD for the data are shown at the top part in each column. 
Colors used for the optical photometry are 
blue for $g$, 
cyan for $V$, 
green for $R_C$ and $r$, and 
red for $I_C$ and $i$. 
Colors used for the NIR photometry are 
blue for $J$, 
green for $H$, and 
red for $K_s$. 
Standard deviations in magnitude unit within a time window are shown at the left-bottom part of the panels. 
Optical PS1 photometry in $gri$-bands and NIR 2MASS photometry in $JHK_s$-bands are shown in dashed lines. 
A thick gray line in the top second panel shows the length of 10~minutes. 
Panels with no data are hatched in gray. 
Large scatters ($\gtrsim0.05$~mag) are partly attributed to bad seeing. 
}\label{fig:fig_lc_m}
\end{figure*}

Magnified views of the light curves 
on these densely observed nights are shown in 
the 1st (optical) and 3rd (NIR) rows of Figure~\ref{fig:fig_lc_m}. 
In this figure, photometry is done for each frame. 
For a comparison, ensemble relative light curves of nearby stars 
with similar brightness on average are also shown 
in the 2nd (optical) and 4th (NIR) rows of the figure. 
As seen in the daily light curves in Figure~\ref{fig:fig_lc}, 
brightness change from night to night is easily seen. 
R.m.s. of the brightness is also shown in the figure. 

Observed intranight variabilities are as small as $0.03-0.11$~mag, depending 
on the filters and epochs (the data quality) 
and no significant intranight variabilities are detected in our dataset. 
The large scatters ($\gtrsim0.05$~mag) seen in some panels of the figure are partly attributed to bad seeing. 
In \citet{sagar2004}, 
a few to 10 percent amplitude ($\lesssim0.1$~mag, $14.1$\% at maximum) 
intranight $R_C$-band variability are seen for 11~blazars (6 BL~Lacs and 
5 radio core-dominated quasars). 
Typical observation duration in a night in \citet{sagar2004} is 6.5~hours 
and the total number of the observed nights is 47. 
They find that a duty cycle of intranight optical variability is $\sim60$\% for BL~Lacs. 
Similarly, \citet{paliya2017} also monitored 17~blazars for $137$~hours 
in $R_C$-band 
and also obtained 
a high duty cycle of $\sim40$\% for blazars. 
Compared with these studies, the total duration (sum) 
of our observations shown in Figure~\ref{fig:fig_lc_m} 
is $\sim18$~hours 
and much shorter by factor of $\sim17$ \citep{sagar2004} and $\sim8$ \citep{paliya2017}. 
Then, it is difficult to say any consistency of our non-detection with 
these two observations. 

\subsubsection{Second-Scale Variability}\label{sec:sec_secondvari}

The fastest variability detected for blazars so far is a few minutes 
in optical \citep{sasada2008} 
and 
$\gamma$-ray 
\citep{albert2007,aharonian2007,vovk2013,vovk2015,ackermann2016}, 
which corresponds to a comparable size of a black hole, 
which indicates a small emitting region in a relativistic jet.
A mass of a black hole hosting a blazar is expected 
to be as massive as $\sim10^{9}$~M$_\odot$ in general \citep{castignani2013}, 
and the black hole mass of \target\ is estimated to be $\sim3\times10^{8}$~M$_\odot$ \citep{padovani2019} 
by assuming the typical host galaxy luminosity of blazars \citep{paiano2018} 
and black hole mass and bulge mass relation \citep{mclure2002}. 
Detection of variability in a shorter time scale would put 
a tight constraint 
on the size of an emitting region
with a usually assumed Doppler boosting factor. 
If second scale variability was detected, this might be attributed to 
an apparent change in viewing angle to a bent relativistic jet \citep{raiteri2017}. 

\begin{figure}
 \begin{center}
  \includegraphics[width=95.4mm,bb=0 0 634 396]{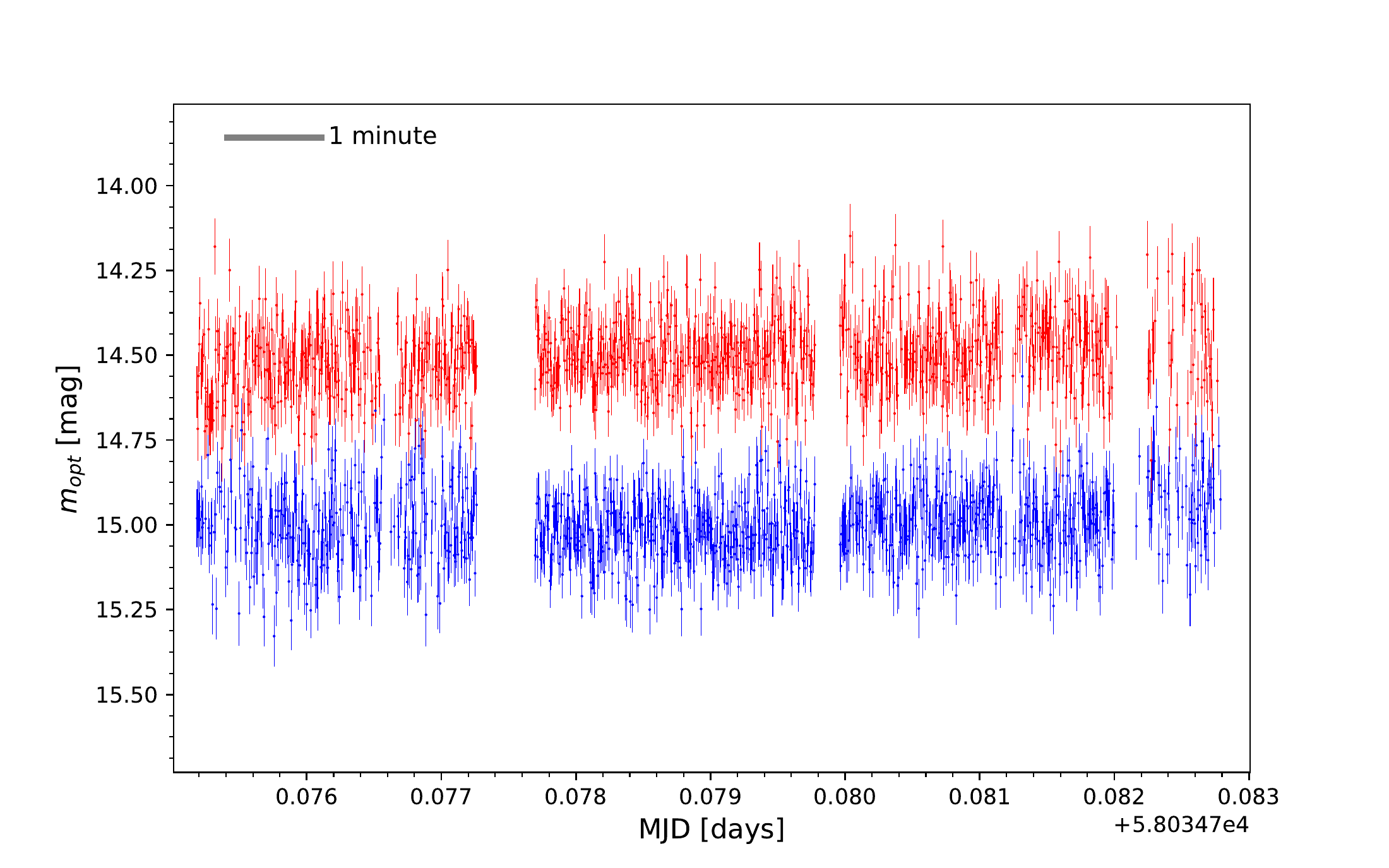}
 \end{center}
\caption{
Optical light curves 
of \target\ (red) and 
a nearby star (blue) 
taken with \tomoe. 
}\label{fig:fig_lc_tomoe}
\end{figure}

Second-scale variability is 
investigated with the 2~fps 
\tomoe\ data. 
Four sets of 180-second (360~frames) exposure are obtained. 
Detection of \target\ is sometimes marginal 
and we use only photometric data of signal-to-noise ratios of $>10\sigma$. 
As a result, the photometric data 
especially 
in the last (fourth) exposure are partly removed in our analysis and shown in Figure~\ref{fig:fig_lc_tomoe}. 
The resultant SF is shown 
in the leftmost part of the top panel of Figure~\ref{fig:fig_sf}.
In our dataset, 
no significant rapid variability in a second time scale is detected. 

\subsection{Optical and NIR colors of \target}\label{sec:sec_sed}

\begin{figure*}
 \begin{center}
  \includegraphics[width=164.5mm,bb=0 0 616 309]{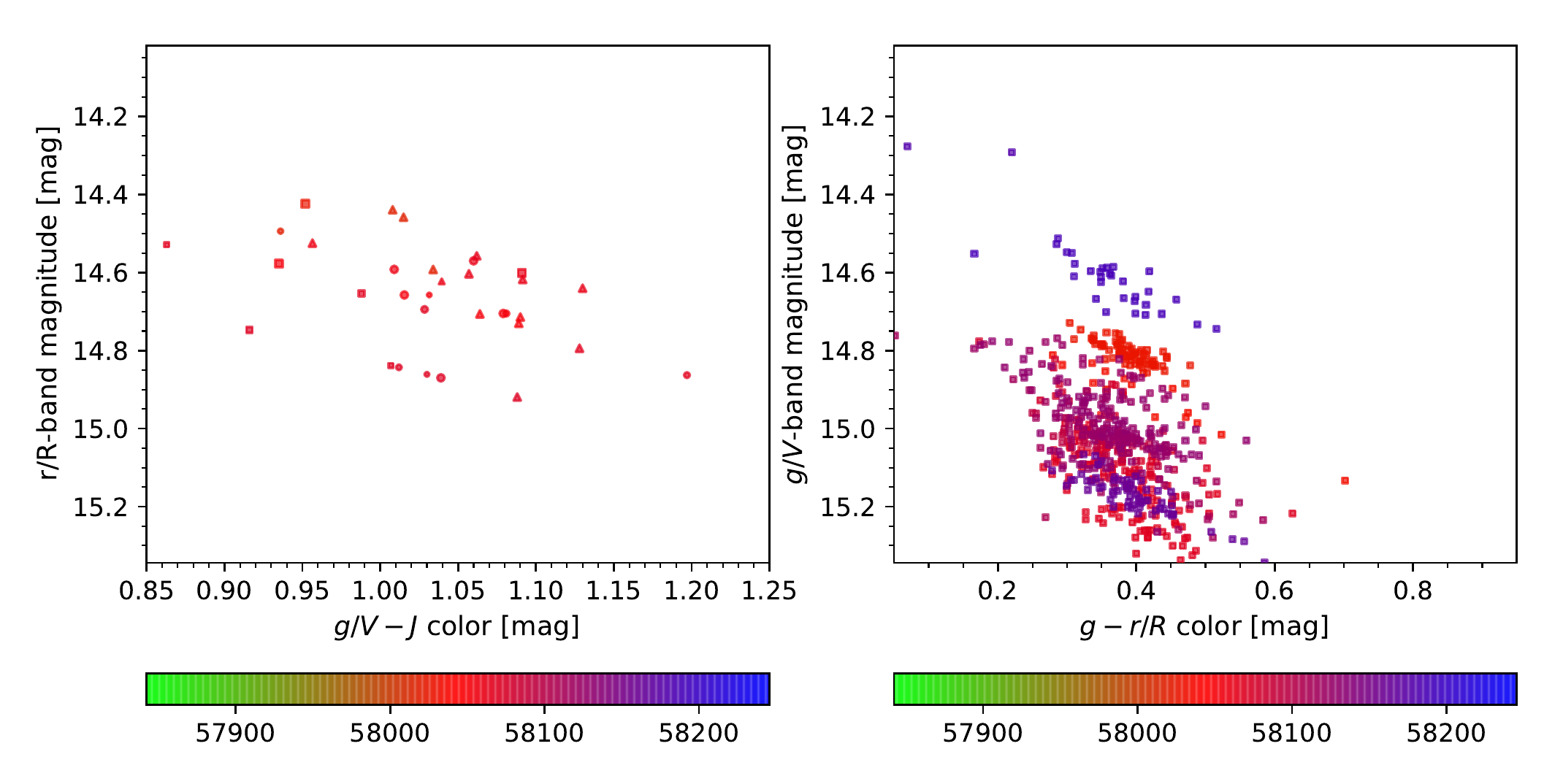}
 \end{center}
\caption{
Optical magnitudes 
($r$ or $R_C$ in the left panel and  
$g$ or $V$ in the right panel) 
as functions of 
optical-NIR color (left panel; $g-J$ or $V-J$)
and optical color (right panel; $g-r$ or $g-R_C$). 
Different symbols indicate data taken with different telescopes/instruments 
in the same way as in Figure~\ref{fig:fig_lc}. 
}\label{fig:fig_optmag2a}
\end{figure*}

Temporal changes of 
optical, optical-NIR, NIR-NIR colors 
are shown 
in 
Figure~\ref{fig:fig_lc} (entire light curves), 
Figure~\ref{fig:fig_lc_m} (intranight light curves),  
and Figure~\ref{fig:fig_optmag2a} (correlation between magnitudes and colors). 

The range of the optical colors of 
$g-r$ and $g-R$ 
is 0.2-0.5~mag. 
A bluer-when-brighter 
trend is seen in the color-magnitude diagrams
(Figure~\ref{fig:fig_optmag2a}).  
This is consistent with an idea that \target\  is a BL~Lac-type blazar 
with little contribution of its accretion disk 
to optical-NIR emission \citep{bonning2012}. 
In Figure~\ref{fig:fig_optmag2a}, 
bluish data points indicating the data obtained 
around MJD=58200 
are offset by $\sim0.4$~mag from the most crowded (redish) data region. 
These bluish data also follow a tighter blue-when-brighter trend. 
In summary, data points in different epochs follow different 
color-magnitude relations. 
This behavior is also observed for other blazars, 
for example, OJ~287, and 
indicates different activity states 
between the different loci in the color-magnitude diagram 
in different epochs \citep{bonning2012}. 
These bluish points are data taken after the $\gamma$-ray flare 
in March 2018 \citep{ojha2018} while the redish points are taken after 
the IceCube neutrino detection \citep{icecube2018_science1}. 
In these two epochs, increased $\gamma$-ray fluxes are detected with Fermi 
but the optical and NIR color-magnitude relations are different from each other. 
In general, 
the brighter locations of the bluish points at a given color may be attributed 
to 
high-energy electron injection into jet-emitting region 
or 
an emergence of much brighter accretion disk than usual 
possibly due to an accretion state change. 
Bluer-when-brighter trends of BL~Lacs are sometimes attributed to a presumption that 
the objects are in high state \citep{zhang2015}. 
\target\ may 
show this trend 
in any state 
considering 
that almost featureless power-law continua are always observed and 
that the equivalent widths (EWs) of the emission lines of \target\
(EW$_{\rm{[OIII]}}=0.17$\AA\ at most; \cite{paiano2018}) are small 
compared to previously measured EWs of blazars 
(although many of these measurements are upper limits) \citep{landt2004}. 

The range of $r-J$ or $R_C-J$ colors of \target\ is 
from 0.9 to 1.2 (Figure~\ref{fig:fig_optmag2a}), 
roughly corresponding to 1.8 to 2.1 in the Vega system. 
The $r-J$ and $R_C-J$ 
color changes are as small as $\sim0.3$~mag. 
Although the previous studies examined different colors 
(e.g., $V-J$ colors in \cite{ikejiri2011}), 
these are typical for ISP blazars. 
The bluer-when-brighter trend is also seen. 
Note that no data points are shown in the panel 
of Figure~\ref{fig:fig_optmag2a} 
after the $\gamma$-ray flare in March 2018 reported by \citet{ojha2018}. 

Intranight changes in optical colors are shown in the bottom panels of Figure~\ref{fig:fig_lc_m}. 
The $g-r$ or $g-R_C$ colors are almost constant, about 0.4~mag all over the nights. 
NIR colors of $J-H$ and $H-K_s$ are also almost constant in time, 0.3-0.4~mag. 
Note that no significant intranight variability in any band is 
detected (\S\ref{sec:sec_intranightvari}). 

\subsection{Polarization of \target}\label{sec:sec_optnirpol_result}

\subsubsection{Temporal Changes of Polarization}\label{sec:sec_optnirpol_result_lc}

Our measurement results of polarization 
in $R_C$ and $J$-bands 
are summarized in Table~\ref{tab:tab_pol2}, 
including the optical polarimetric measurement taken with 
RINGO3 on the 2-m Liverpool telescope \citep{steele2018}. 
The temporal changes of the polarization degrees and angles are shown 
in Figure~\ref{fig:fig_lc}. 

\begin{table*}[!htbp]
  \tbl{Polarization measurements of \target\ with Kanata/HONIR.}{
  \begin{tabular}{ccrrrrl}
      \hline
      Date (UT) & MJD & \multicolumn{2}{c}{Polarization Degree} & \multicolumn{2}{c}{Polarization Angle} & Note\\
                &     & $R_C$ & $J$ & $R_C$ & $J$ & \\
      \hline
      2017-09-30 & 58026.771 & $7.61\pm0.49$ & - & $24.78 \pm 2.65$ & - \\ 2017-10-26 & 58052.770 & $2.42\pm0.39$ & - & $17.71 \pm 5.54$ & - \\ 2017-10-29 & 58055.736 & $6.87\pm1.09$ & - & $10.05 \pm 4.94$ & - \\ 2017-11-09 & 58066.790 & $4.32\pm0.36$ & $4.47\pm0.78$ & $16.19 \pm 2.47$ & $19.70 \pm 2.66$ \\ 2017-11-16 & 58073.761 & $6.21\pm2.92$ & - & $30.11 \pm 2.37$ & - \\ 2018-03-14 & (*)58192     & $\sim14$         & - & -                  & - & $R$, $I$ bands, \citep{steele2018}\\ 2018-03-22 & 58199.463 & $1.50\pm0.67$ & $3.13\pm1.22$ & $84.23 \pm 12.91$ & $65.33 \pm 14.21$ \\ 2018-03-24 & 58201.468 & $5.90\pm1.79$ & - & $54.72 \pm 9.41$ & - \\ 2018-03-25 & 58202.466 & $6.59\pm2.07$ & - & $49.79 \pm 11.90$ & - \\ 2018-03-26 & 58203.454 & $7.20\pm1.47$ & $7.47\pm0.61$ & $37.60 \pm 2.15$ & $45.62 \pm 3.28$ \\ 2018-03-27 & 58204.462 & $8.01\pm2.30$ & - & $27.21 \pm 2.87$ & - \\ 2018-03-28 & 58205.451 & $8.58\pm0.53$ & $7.76\pm1.52$ & $21.56 \pm 1.33$ & $35.36 \pm 2.88$ \\ 2018-03-31 & 58208.469 & $10.15\pm1.33$ & $11.84\pm1.75$ & $47.04 \pm 1.90$ & $45.86 \pm 3.29$ \\ 2018-04-02 & 58210.451 & $16.99\pm0.47$ & - & $40.55 \pm 0.35$ & - \\ 2018-04-03 & 58211.464 & $18.88\pm0.76$ & $16.27\pm0.78$ & $46.17 \pm 2.13$ & $43.91 \pm 1.65$ \\ 2018-04-08 & 58216.451 & $19.26\pm1.36$ & $20.49\pm4.59$ & $29.69 \pm 1.90$ & $32.63 \pm 2.48$ \\ 2018-04-09 & 58217.464 & $10.47\pm0.40$ & $8.80\pm1.33$ & $40.74 \pm 3.88$ & $31.69 \pm 1.46$ \\ 2018-04-13 & 58221.469 & $7.23\pm0.51$ & - & $56.48 \pm 2.01$ & - \\ 
	\hline
    \end{tabular}}
    \label{tab:tab_pol2}
\begin{tabnote}
(*) is calculated by assuming the observations is done at UT=0 hours. 
\end{tabnote}
\end{table*}

For the first 5~data points taken within 1.5~months 
after the alert (defined as ``first epoch''), 
the polarization degrees are as small as 2--8\%\ 
as partly reported in \citet{yamanaka2017_IC170922a}. 
About 6~months after the neutrino detection (defined as ``second epoch''), 
\citet{ojha2018} reported that 
a flare of the highest daily averaged $\gamma$-ray flux for \target\ 
was detected with {\it Fermi} on March 13, 2018. 
Soon after that, 
\citet{steele2018} carried out 
polarimetric observation on the night of March~14, 2018
with Liverpool/RINGO3 and 
found that optical polarization degree increases up to $\sim14$\% 
at wavelengths roughly corresponding to 
the $R_C$ and $I_C$-bands.
As shown in Figure~\ref{fig:fig_lc} and Table~\ref{tab:tab_pol2}, 
our subsequent observations with Kanata/HONIR 
(12 polarization measurements for 23~days starting 8~days after the RINGO3 observation) 
indicate that 
optical polarization degree again decreases down to 
1.5\%\ on March 22, 2018, 
which is even lower than the observed level in the first epoch. 
After this decrease, the polarization degrees gradually 
increase up to $\sim20$\% during about two weeks 
and decreases again down to 7.2\%, which is as low as those in the first epoch. 
Throughout this period, the $J$-band polarization exhibits a
similar time variation behavior to that of 
$R_C$-band. 

In ISPs, 
typical polarization degrees and their temporal change 
are 
$\sim30$~percent and 
$\sim20$~percent or less, respectively, among the samples observed by 
\citet{ikejiri2011}, \citet{itoh2016}, and \citet{jermak2016}. 
The polarization degrees and its temporal change observed in \target\  is comparable
to or less than them, being consistent with those objects.

The polarization angles are roughly
constant at $\sim20^{\circ}\pm10^{\circ}$ 
in both $R_C$ and $J$-bands 
in first epoch. 
In the second epoch, our HONIR measurements indicate 
that the polarization angles are 20--90~deg and significantly different from 
those in the first epoch. 
Following the polarization degree increase from 1.5\%\ to $\sim20$\%, 
the polarization angles first 
change from 84~deg to 22~deg 
and back to 30-60~deg. 
The position angles in $J$-band also show a similar behavior 
to $R_C$-band. 

\subsubsection{Correlations between Optical Brightness and Polarization}\label{sec:sec_optnirpol_result_corr}

\begin{figure}[!htbp]
 \begin{center}
  \includegraphics[width=75.5mm,bb=0 0 396 734]{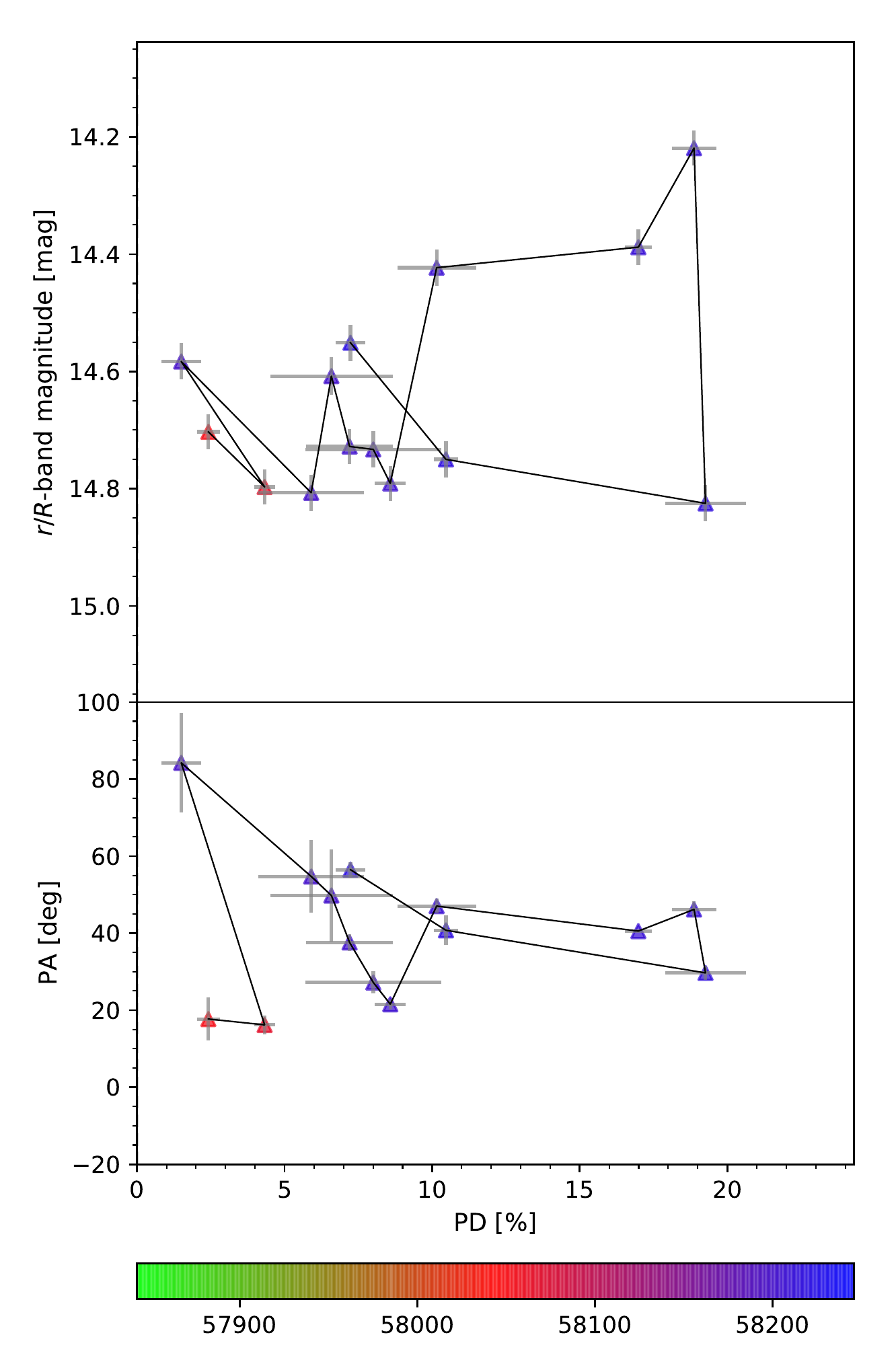}
 \end{center}
\caption{
$r$ or $R_C$-band magnitude 
and 
$R_C$-band polarization angle 
as a function of 
$R_C$-band polarization degree. 
All the data points are connected in time sequence. 
Different symbols indicate data taken with different telescopes/instruments 
in the same way as in Figure~\ref{fig:fig_lc}. 
}\label{fig:fig_optmag2bpol}
\end{figure}

The top panel of Figure~\ref{fig:fig_optmag2bpol} shows 
a relation between the polarization degrees and 
optical $r$ or $R_C$-band magnitudes. 
Although 
optical brightness and polarization degrees look roughly 
coincident with each other as a whole (Figure~\ref{fig:fig_lc}), 
almost no correlations are seen 
during the rapid changes in the second epoch 
as indicated by bluish points in Figure~\ref{fig:fig_optmag2bpol}. 
This is discussed later in this subsection.

\begin{figure}[!htbp]
 \begin{center}
  \includegraphics[width=83.5mm,bb=0 0 382 508]{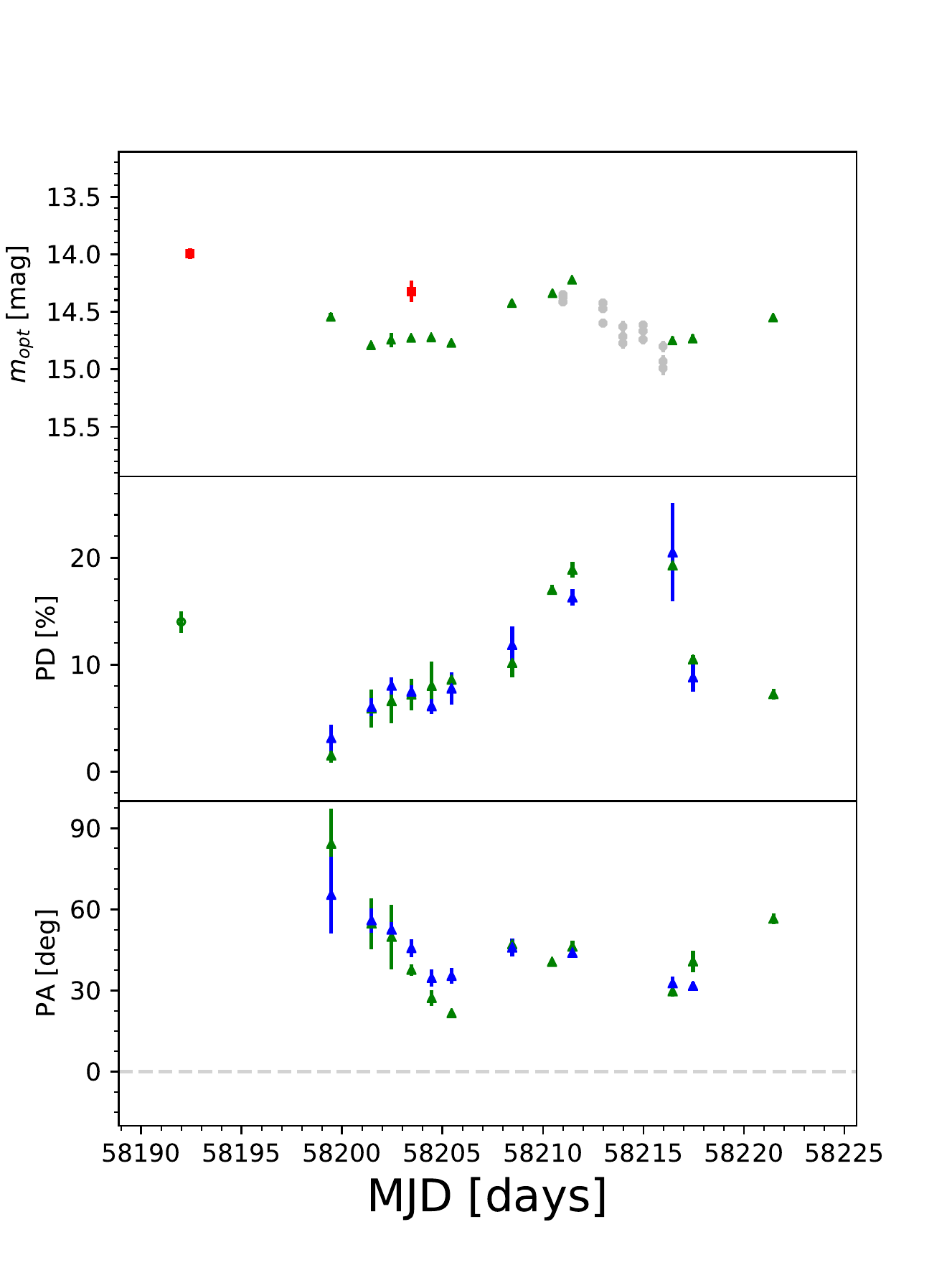}
 \end{center}
\caption{
A magnified view of 
temporal changes of optical magnitudes, 
polarization degrees, and 
polarization angles 
around the $\gamma$-ray flare reported by \citet{ojha2018}. 
The symbols and colors are the same as those in Figure~\ref{fig:fig_lc}. 
$r$,$R_C$-band in green, 
$i$,$I_C$-band in red, 
and 
$J$-band in blue. 
}\label{fig:fig_lc2ndepoch}
\end{figure}

Polarization changes in optical wavelengths associated $\gamma$-ray flares 
are reported in previous literature \citep{sasada2008,abdo2010_3c279}. 
In the second epoch, the polarization change of \target\ 
seems associated to the $\gamma$-ray flare \citep{ojha2018} 
although such a correlated behavior is not clear in the first epoch 
because the polarization measurements are not done right after (or before) the neutrino detection, 
the measurements are sparse, 
and the number of the measurements is small. 
In \citet{itoh2013}, 
no brightness flare was observed during their first polarization flare 
for the famous blazar, CTA~102 (an FSRQ at $z=1.037$), 
indicating that the polarization degree does not necessarily correlate with brightness. 

The lack of strong correlations between optical brightness and polarization degree in our data for \target\ 
is similarly observed in previous literature for other blazars \citep{ikejiri2011,jermak2016},  
although 
a significant 
positive correlation 
between ``amplitudes'' of flux and polarization degree 
is detected 
for two blazars (AO~0235+164 and PKS~1510-089) 
in 10-day time scale \citep{sasada2011}. 
These observed weak correlations 
could be partly due to ignorance of a possible time lag between 
temporal changes of fluxes and polarization degrees \citep{uemura2017}. 
Figure~\ref{fig:fig_lc2ndepoch} is a magnified view of Figure~\ref{fig:fig_lc} around the second epoch. 
The peak of polarization degrees is around MJD$\sim$58211--58217~days 
while that of optical brightness is around MJD$\sim$58210--58211~days. 
This indicates that the optical brightness change proceeds the polarization degree change, 
which is 
the opposite sense to that observed for a BL~Lac, PKS~1749+096, reported by \citet{uemura2017}. 
This lag partly makes the correlation worse in Figure~\ref{fig:fig_optmag2bpol}. 
A positive correlation would be seen 
if the data point in MJD=58216.5 (the right bottom point 
in Figure~\ref{fig:fig_optmag2bpol}) is ignored. 

The polarization angles also change in time by $\sim70$~deg in the second epoch 
although the change is not so drastically large. 
Figure~\ref{fig:fig_lc2ndepoch} indicates that 
the polarization angles decrease as the polarization degrees increase 
in the period of MJD$=58198-58207$. 
The polarization degrees still increase after that, however, 
the polarization angles do not show a systematic decrease. 
These make the correlation between the polarization angles and degrees poor 
as shown in the bottom panel of Figure~\ref{fig:fig_optmag2bpol}. 
Changes of polarization angles are observed for many blazars \citep{itoh2016,hovatta2016}. 
The observed change of the polarization angles for \target\ is not so large, which is 
sometimes explained by 
a curved structure of a relativistic jet 
\citep{abdo2010_3c279,sorcia2014}. 

\subsection{Spectra of \target}\label{sec:sec_spec}

\begin{figure}
 \begin{center}
  \includegraphics[width=90mm,bb=0 0 408 624]{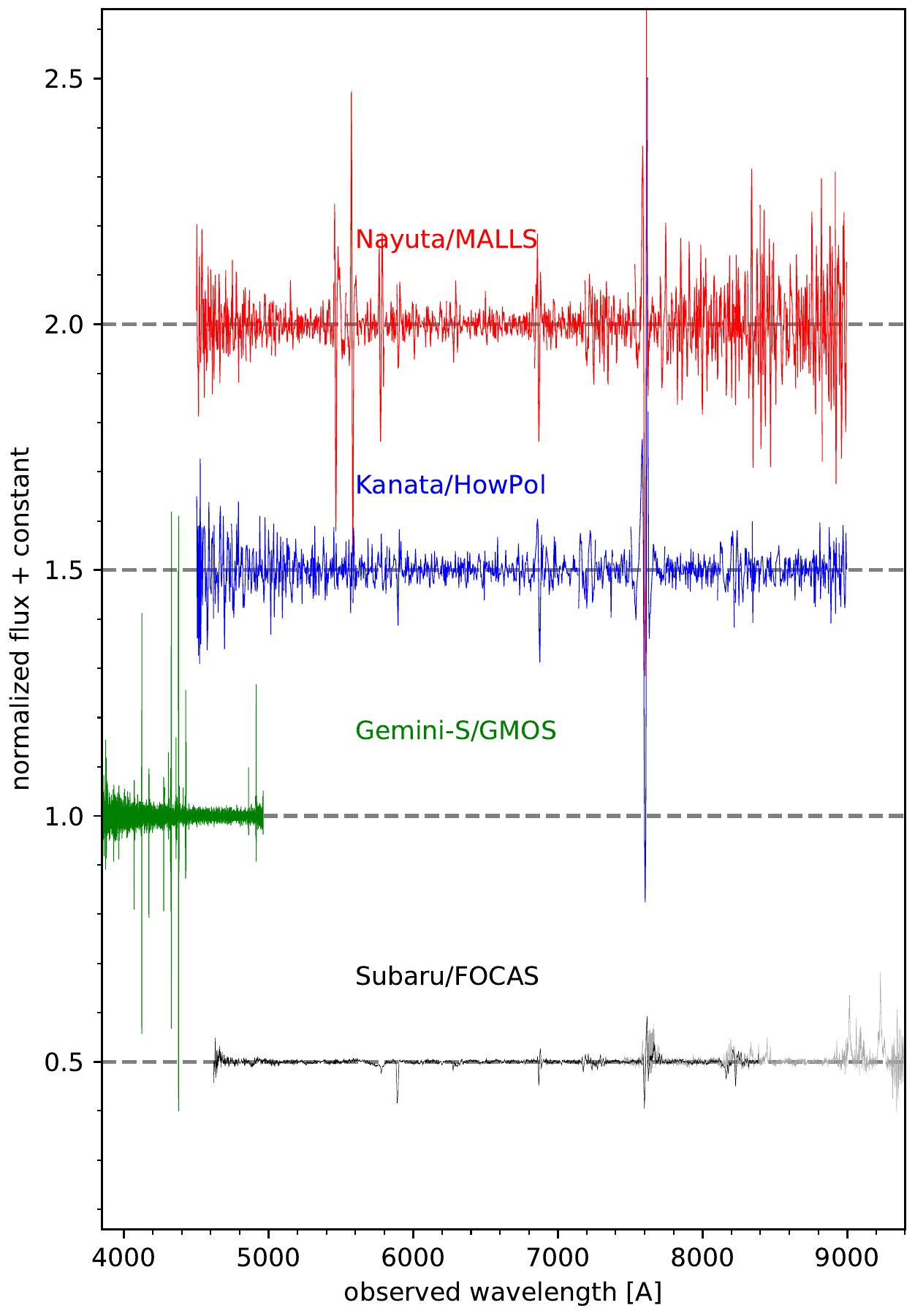}
 \end{center}
\caption{
Optical spectra of \target\ taken with 
Nayuta/MALLS (red), 
Kanata/HOWPol (blue), 
Gemini-South/GMOS (green), 
and 
Subaru/FOCAS (black and gray). 
The Subaru/FOCAS spectra are the same as those shown in \citet{icecube2018_science1}. 
There are no changes in spectral features. 
Noisy wavelength ranges are omitted. 
All the spiky features seen in the spectra are all atmospheric. 
}\label{fig:fig_spec}
\end{figure}

All of our three new spectra of \target\ are shown in Figure~\ref{fig:fig_spec}. 
The Subaru/FOCAS spectra in the two different setups shown in \citet{icecube2018_science1} 
are also plotted as references. 
All these spectra were taken roughly in similar epochs, 
$\sim1$~week to $\sim1.5$~months 
after the neutrino detection, and 
before the GTC/OSIRIS spectrum 
which conclusively determines the redshift of \target\ was taken \citep{paiano2018}. 
In the epochs of our observations, 
\target\ is slightly brighter ($g=14.8-15.1$)
than that in the GTC observation ($g=15.4$; \cite{paiano2018}), 
which makes line detections more difficult. 
All of our three spectra show basically featureless continua and 
no significant emission or absorption lines are detected, 
except for the weak emission line in the Subaru/FOCAS spectrum \citep{icecube2018_science1}. 
These are quantitatively 
consistent with the spectrum of \citet{paiano2018}. 
Any significant changes between the different epochs are not detected. 

\section{Summary}\label{sec:sec_summary}

We first made optical and NIR imaging observations to search for a candidate of 
the neutrino source of \thisevent.  
We found that \target\ was rapidly fading in NIR 
in a day scale. 
Motivated by this discovery of the rapid NIR variability, 
the $\gamma$-ray flare was discovered in the Fermi/LAT monitoring data. 
We conducted monitoring observations of \target\ with 
Akeno/MITSuME, 
Kiso/KWFC, 
Kiso/\tomoe, 
Kyoto 0.4-m, 
Kanata/HONIR, and 
Subaru/HSC. 
Polarization imaging data were also taken with Kanata/HONIR at 17~epochs. 
We also took optical spectra \sout{four times} five times with 
Kanata/HOWPol, Nayuta/MALLS, Gemini-N/GMOS, and Subaru/FOCAS. 

Combined with ASAS-SN optical monitoring data and Fermi/LAT $\gamma$-ray data in addition to our data described above, we find 
\begin{itemize}
    \item daily variability is significant and the amplitude is as large as \varidaily~mag, 
    \item no significant intranight variability is detected, 
    \item no special optical/NIR variability behavior over the observed period, 
    but a very marginal sign of a larger variability in a time scale of $\sim10$ days about 2~months before the neutrino detection, 
    \item a weak correlation between optical/NIR and $\gamma$-ray fluxes
    \item large changes in polarization degrees and angles about 180~days after the neutrino detection,
    \item a weak or no correlation between polarization degree and optical fluxes while some correlated behavior can be seen 
    in a part of the March 2018 ($\sim180$~days after the neutrino detection) data,  
    \item no significant optical spectral changes over the three months after the neutrino detection.
\end{itemize}

In summary, our data do not indicate that 
\target\ is a special blazar (BL~Lac) among other blazars in terms of intranight, daily, 
monthly optical/NIR variability, and optical polarization. 
The neutrino detection is also not a special timing. 

For future Icecube neutrino events, to further examine possible relations 
between high-energy neutrinos and blazar 
as well as blazar variability, more complete blazar catalogs like a recently 
developed one, BROS \citep{itoh2018bros}, 
are required. 
In addition, in optical and NIR wavelengths, 
routine high-cadence imaging and polarization monitoring of blazars are desired. 
Transient survey 
observations are done 
in several wide-field surveys such as Zwicky Transient Facility (ZTF; \cite{graham2019}) and 
Tomo-e Gozen \citep{sako2018}. 
Polarization monitoring is a more expensive observation program but is expected 
to make unique science outputs 
(e.g., \cite{sakimoto2012,itoh2016}). 

\begin{ack}

We thank an anonymous referee for his/her comments to improve the manuscript. 
Observations with the Kanata, Nayuta, MITSuME, IRSF, and Kiso Schmidt telescopes were supported by the Optical and Near-infrared Astronomy Inter-University Cooperation Program and the Grants-in-Aid of the Ministry of Education, Science, Culture, and Sport 
JP23740143, 
JP25800103, 
JP16H02158, 
JP18H04575, 
JP15H02075, 
JP16H06341, 
JP18H05223, 
JP17K14253, 
JP17H04830, 
JP26800103, 
JP19H00693, 
JP17H06363, 
JP17H06362, 
and 
JP24103003. 

This work was partly supported by the joint research program of the
Institute for Cosmic Ray Research (ICRR).
This work was also based in part on data collected at Subaru Telescope, 
which is operated by the National Astronomical Observatory of Japan. 
Based on observations obtained at the international Gemini Observatory, a program of NOIRLab, which is managed by the Association of Universities for Research in Astronomy (AURA) under a cooperative agreement with the National Science Foundation. on behalf of the Gemini Observatory partnership: the National Science Foundation (United States), National Research Council (Canada), Agencia Nacional de Investigaci\'{o}n y Desarrollo (Chile), Ministerio de Ciencia, Tecnolog\'{i}a e Innovaci\'{o}n (Argentina), Minist\'{e}rio da Ci\^{e}ncia, Tecnologia, Inova\c{c}\~{o}es e Comunica\c{c}\~{o}es (Brazil), and Korea Astronomy and Space Science Institute (Republic of Korea).

The IRSF project is a collaboration between Nagoya University and the
SAAO supported by the Grants-in-Aid for Scientific Research on
Priority Areas (A) (No.~10147207 and No.~10147214) and Optical \& 
Near-Infrared Astronomy Inter-University Cooperation Program, from the
Ministry of Education, Culture, Sports, Science and Technology (MEXT)
of Japan and the National Research Foundation (NRF) of South Africa.

The Pan-STARRS1 Surveys (PS1) and the PS1 public science archive have been made possible through contributions by the Institute for Astronomy, the University of Hawaii, the Pan-STARRS Project Office, the Max-Planck Society and its participating institutes, the Max Planck Institute for Astronomy, Heidelberg and the Max Planck Institute for Extraterrestrial Physics, Garching, The Johns Hopkins University, Durham University, the University of Edinburgh, the Queen's University Belfast, the Harvard-Smithsonian Center for Astrophysics, the Las Cumbres Observatory Global Telescope Network Incorporated, the National Central University of Taiwan, the Space Telescope Science Institute, the National Aeronautics and Space Administration under Grant No. NNX08AR22G issued through the Planetary Science Division of the NASA Science Mission Directorate, the National Science Foundation Grant No. AST-1238877, the University of Maryland, Eotvos Lorand University (ELTE), the Los Alamos National Laboratory, and the Gordon and Betty Moore Foundation.

This publication makes use of data products from the Two Micron All Sky Survey, which is a joint project of the University of Massachusetts and the Infrared Processing and Analysis Center/California Institute of Technology, funded by the National Aeronautics and Space Administration and the National Science Foundation.

The data is partly processed using the Gemini IRAF package. 
The authors acknowledge Dr.~N.~Matsunaga for a software to conduct comparisons between our image coordinates and the catalogue values.

\end{ack}

\bibliography{IceCube170922A_OptNIR}

\end{document}